\documentclass[aps,prc,twocolumn,times,graphicx,tighten,showpacs]{revtex4}
\usepackage{bm}
\usepackage{amsmath}
\usepackage{graphicx}
\usepackage{graphics}
\usepackage[dvips]{color}

\begin{document}

\title{Meson-exchange currents and quasielastic neutrino cross sections in the
SuperScaling Approximation model}
\author{J.E. Amaro}
\affiliation{Departamento de F\'{\i}sica At\'{o}mica, Molecular y Nuclear,
Universidad de Granada,
  18071 Granada, SPAIN}
\author{M.B. Barbaro}
\affiliation{Dipartimento di Fisica Teorica, Universit\`a di Torino and
  INFN, Sezione di Torino, Via P. Giuria 1, 10125 Torino, ITALY}
\author{J.A. Caballero}
\affiliation{Departamento de F\'{\i}sica At\'{o}mica, Molecular y Nuclear,
Universidad de Sevilla,
  41080 Sevilla, SPAIN}
\author{T.W. Donnelly}
\affiliation{Center for Theoretical Physics, Laboratory for Nuclear
  Science and Department of Physics, Massachusetts Institute of Technology,
  Cambridge, MA 02139, USA}
\author{C.F. Williamson}
\affiliation{Center for Theoretical Physics, Laboratory for Nuclear
  Science and Department of Physics, Massachusetts Institute of Technology,
  Cambridge, MA 02139, USA}

\begin{abstract}
We evaluate the quasielastic double differential neutrino cross
sections obtained in a phenomenological model based on the superscaling behavior
of electron scattering data.
We compare our results with the recent experimental data for neutrinos 
of MiniBooNE and estimate the contribution of the vector meson-exchange
currents in the 2p-2h sector.
\end{abstract}


\maketitle

\section{Introduction}

Muon neutrino charged current quasielastic (CCQE) double
differential cross sections have recently been measured for the
first time by the MiniBooNE
collaboration~\cite{AguilarArevalo:2010zc}. The results demonstrate
the inadequacy of the Relativistic Fermi Gas (RFG) as a model for
the nuclear system. Indeed the RFG, presently used in the
experimental analysis, underestimates the total cross section,
unless an unusually large {\it ad hoc} value $M_A=1.35$ GeV/c$^2$ is
used for the nucleon axial mass.

The RFG model has the merit of treating properly the relativistic
aspects of the problem. These cannot be neglected for the kinematics
of MiniBooNE, where the neutrino energy reaches values as high as 3
GeV. However, the RFG is clearly too crude to account for the
nuclear dynamics, as is well known from comparisons with electron
scattering data. With this motivation several more sophisticated
relativistic nuclear models have been applied in recent years to
neutrino
reactions~\cite{Barbaro:1996vd,Alberico:1997vh,Meucci:2003cv,Amaro:2005dn,Meucci:2006ir,Caballero:2005sj,Amaro:2006if,Antonov:2006md,Amaro:2006pr,Amaro:2006tf,Caballero:2006wi,Caballero:2007tz,Ivanov:2008ng}.
The comparison with the new experimental
data~\cite{AguilarArevalo:2010zc} allows for a detailed test of the
corresponding predictions.

Beyond the above-mentioned microscopic relativistic models, a
phenomenological ``SuperScaling'' approach (indicated in what
follows as ``SuSA'') has been proposed in \cite{Amaro:2004bs}, based
on the assumed universality of the scaling function for
electromagnetic and weak interactions. Analyses of inclusive $(e,e^\prime)$ 
data have demonstrated that at energy transfers below the quasielastic 
(QE) peak superscaling is fulfilled rather
well~\cite{Day:1990mf,Donnelly:1998xg,Donnelly:1999sw} 
(see also \cite{Jourdan:1996ut}): this means
that the reduced cross section is largely independent of the
momentum transfer (first-kind scaling) and nuclear target
(second-kind scaling), when represented as a function of the
appropriate scaling variable. From these analyses a phenomenological
scaling function has been extracted from the longitudinal QE electron scattering response and used
to predict neutrino-nucleus cross sections by assuming that this single universal scaling function is appropriate 
for all of the various responses involved (CC, CL, LL, T(VV), T(AA) and T$^{\prime}$(VA); see~\cite{Amaro:2004bs}) and multiplying it by the corresponding elementary weak cross sections.

Although being far more realistic than the RFG, the superscaling
approach described above is based on some assumptions.
First, in studies of inclusive QE electron scattering it assumes the equality of the longitudinal and transverse
scaling functions. This property, which is known as scaling of the
zeroth kind, has been tested in various models and shown to be violated to some extent: 
for example Relativistic Mean Field (RMF) theory
yields a transverse scaling function which is typically 20\% or so larger than
the longitudinal one~\cite{Caballero:2006wi,Caballero:2007tz}. In fact, this is exactly what is observed 
when one examines the existing L/T separated data. Once the effects that are expected to break 
scaling of the zeroth kind are removed, namely, inelastic contributions and effects stemming from 
two-particle-emission meson-exchange currents (see below) which are predominantly transverse in nature, one finds that the remaining transverse scaling function is clearly larger than the longitudinal one. Thus, when proceeding to studies of CCQE cross sections, one should also expect to have some violations of scaling of the zeroth kind as well.
Second, the charged-current neutrino responses are purely isovector, whereas the
electromagnetic ones contain both isoscalar and isovector
components and the former involve axial-vector as well as vector responses. One then has to invoke a further kind of scaling, namely
the independence of the scaling function of the choice of isospin
channel --- so-called scaling of the third kind. The interplay of scaling from the various contributions was first explored in \cite{Caballero:2007tz}. 
Finally, and most important, at energies above the QE peak scaling
is violated in the transverse channel by effects which go beyond the
impulse approximation: inelastic scattering, meson-exchange currents
(MEC) and the associated correlations which must be considered
together with the MEC in order to conserve the electromagnetic current.

In this paper we evaluate the double differential CCQE cross sections
in the SuSA approach and discuss the impact of meson-exchange currents
on the process. We are motivated by the fact that modeling at the level of the impulse
approximation (as is the case for the RFG or for the spectral function approach of \cite{Benhar:2005dj,Ben10,Jus10})
under-predicts the measured CCQE cross sections and seems to call for a significant modification of the axial mass. However, more sophisticated approaches than the RFG such as SuSA and the other modeling discussed below are available and the situation may not be quite so simple. For instance, previous non-relativistic calculations~\cite{Mar09,Mar10} indicate that the 2p-2h
excitations may be able to account for the large measured CCQE cross
section, although a comparison of 2p-2h contributions with the MiniBooNE
data is not yet available.  However, it should be emphasized that the
kinematical regions explored under the integral over the neutrino flux
extends over relativistic domains, and a relativistic treatment of
the nuclear excitations is needed; this has clearly been shown to be necessary for electron scattering. 
In contrast, the 2p-2h MEC considered in the present work are taken from the fully relativistic model
of \cite{De Pace:2003xu}, where it was shown that relativistic
effects are important to describe the nuclear transverse response
function for momentum transfers above 500 MeV/c.


\section{Double differential CCQE cross sections in the SuSA model}
\begin{figure*}[ht]
\label{fig:cosRFGSUSA}
\includegraphics[scale=0.35]{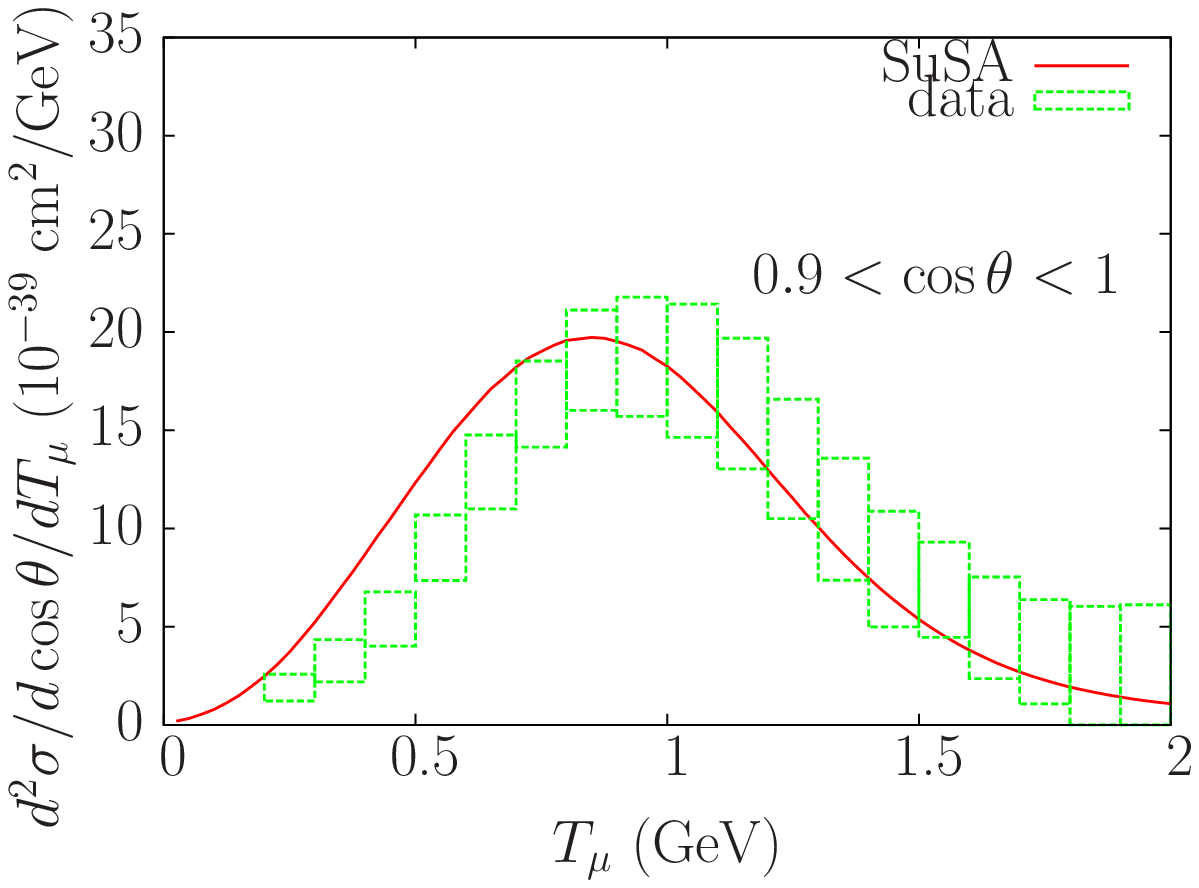}%
\includegraphics[scale=0.35]{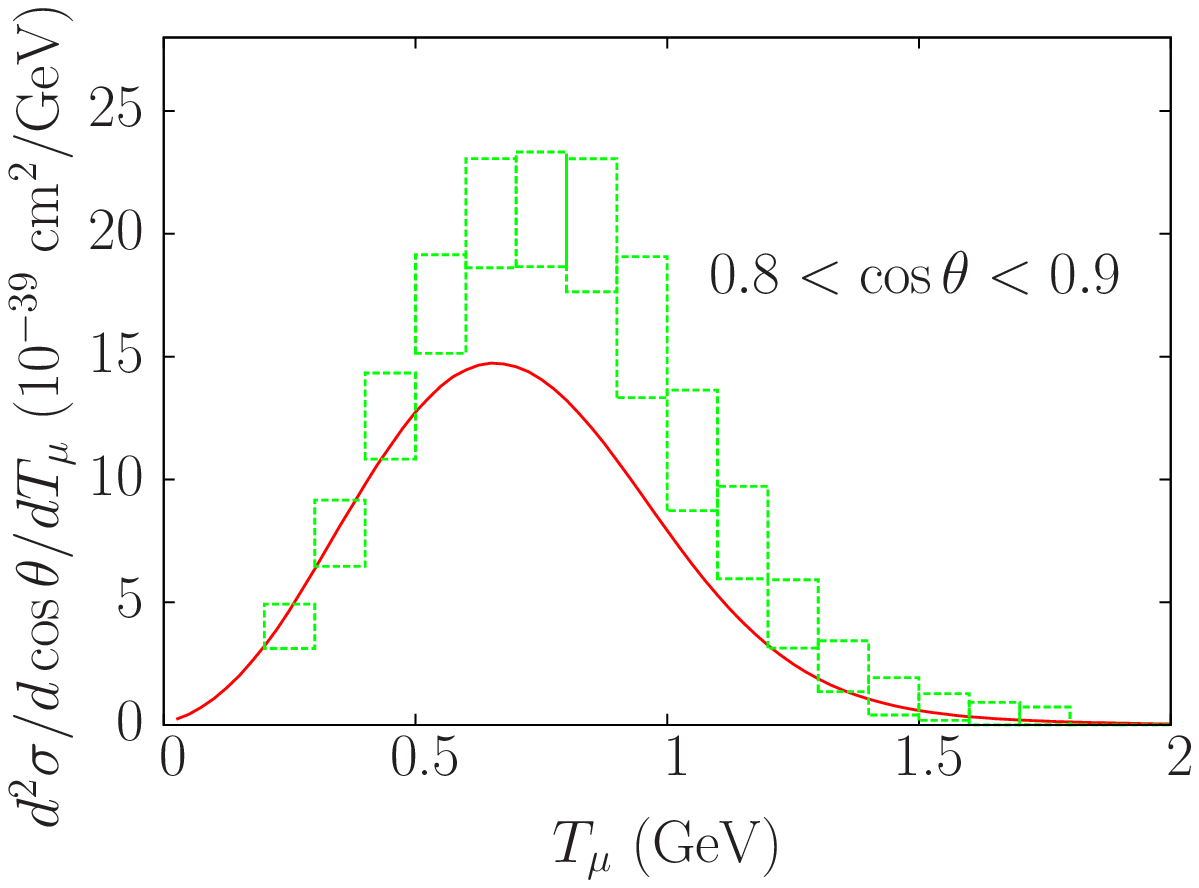}%
\includegraphics[scale=0.35]{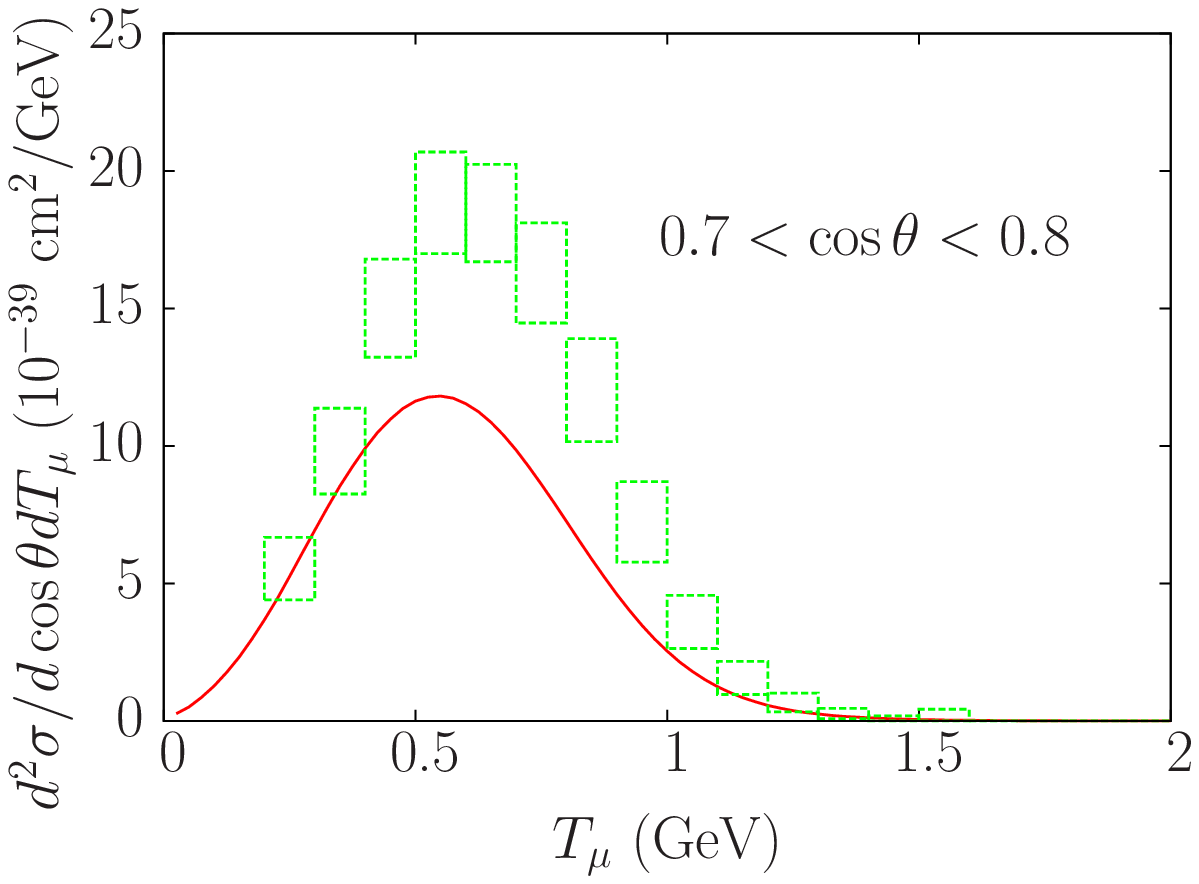}%
\includegraphics[scale=0.35]{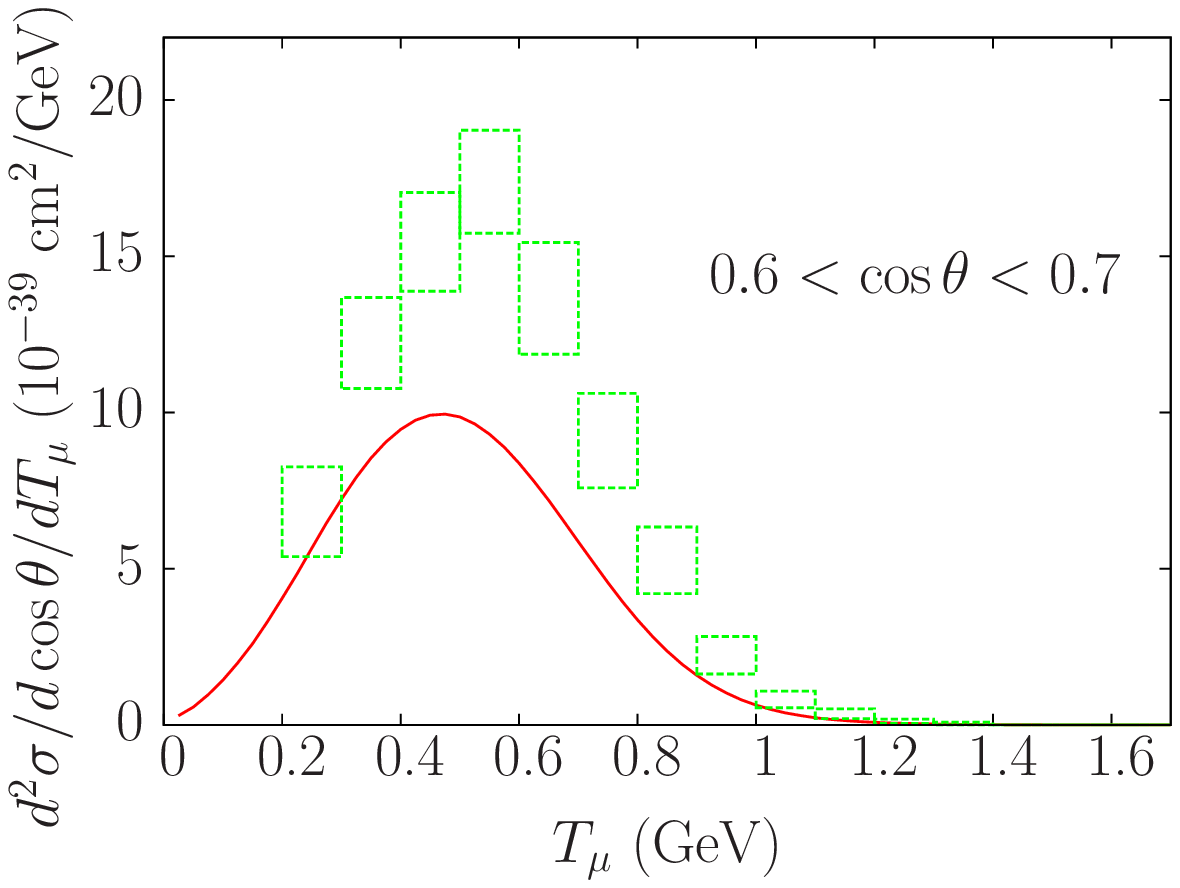}%
\\
\includegraphics[scale=0.35]{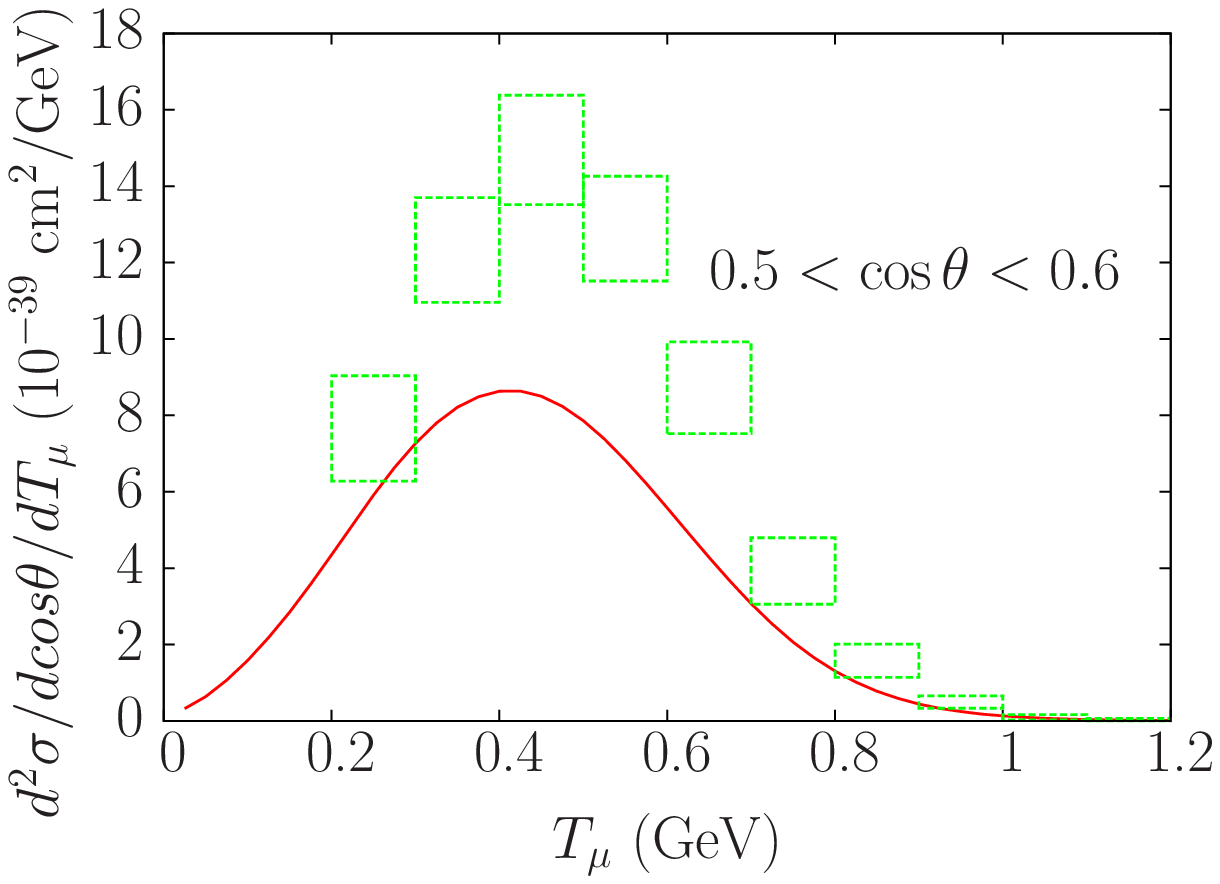}%
\includegraphics[scale=0.35]{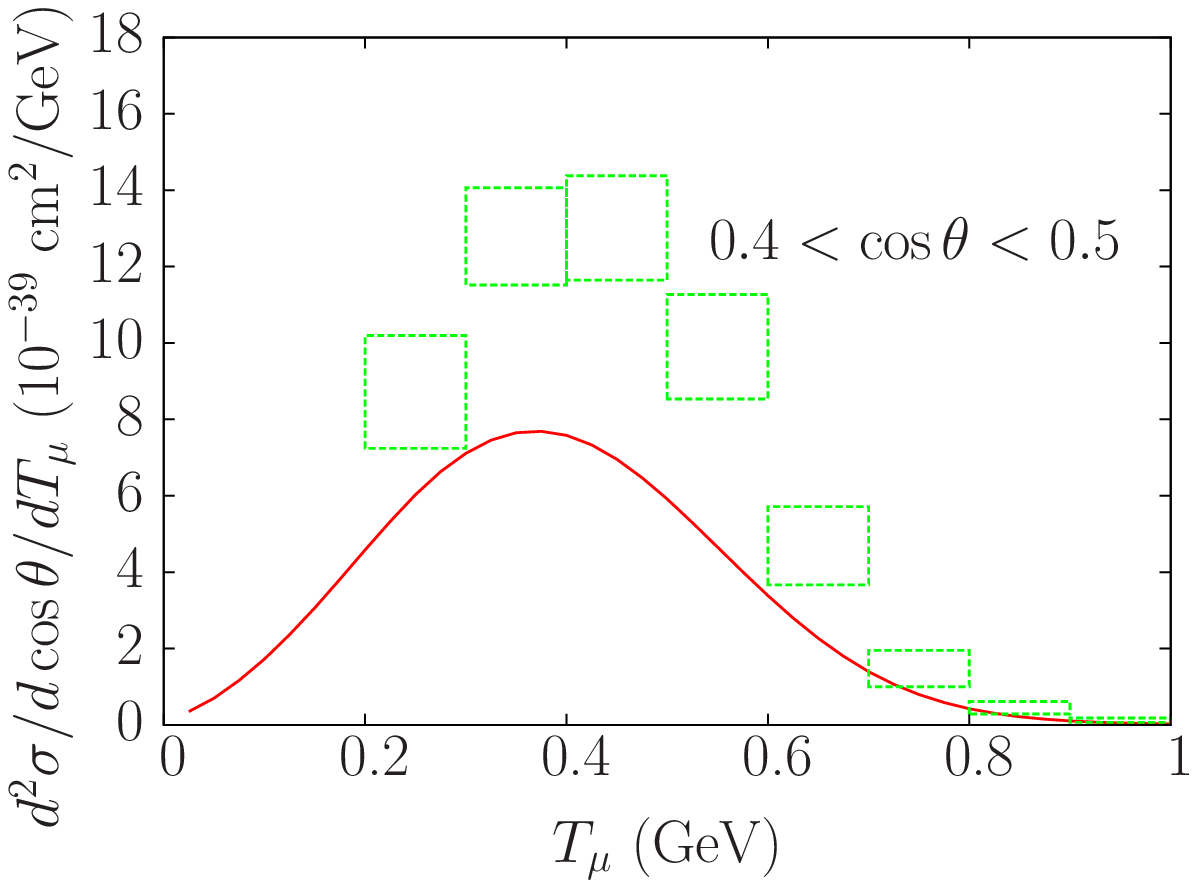}%
\includegraphics[scale=0.35]{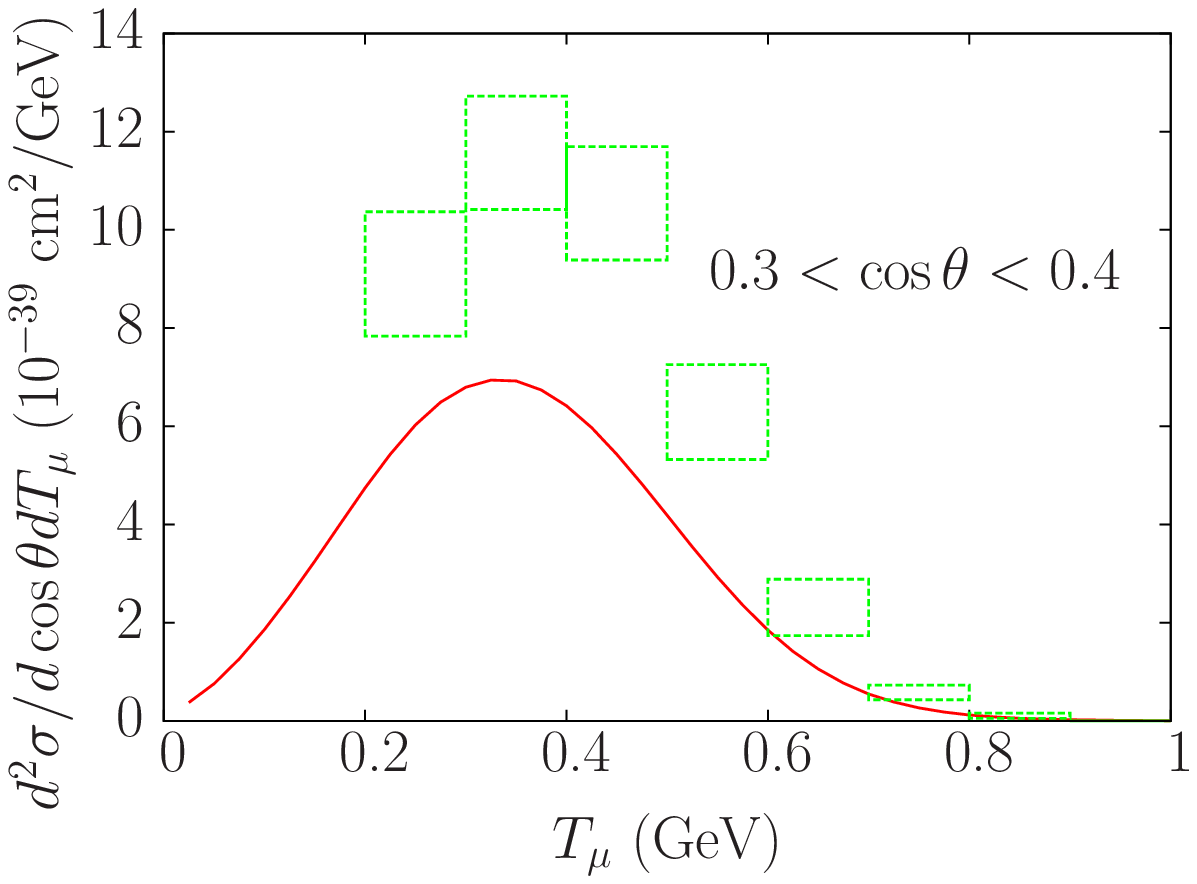}%
\includegraphics[scale=0.35]{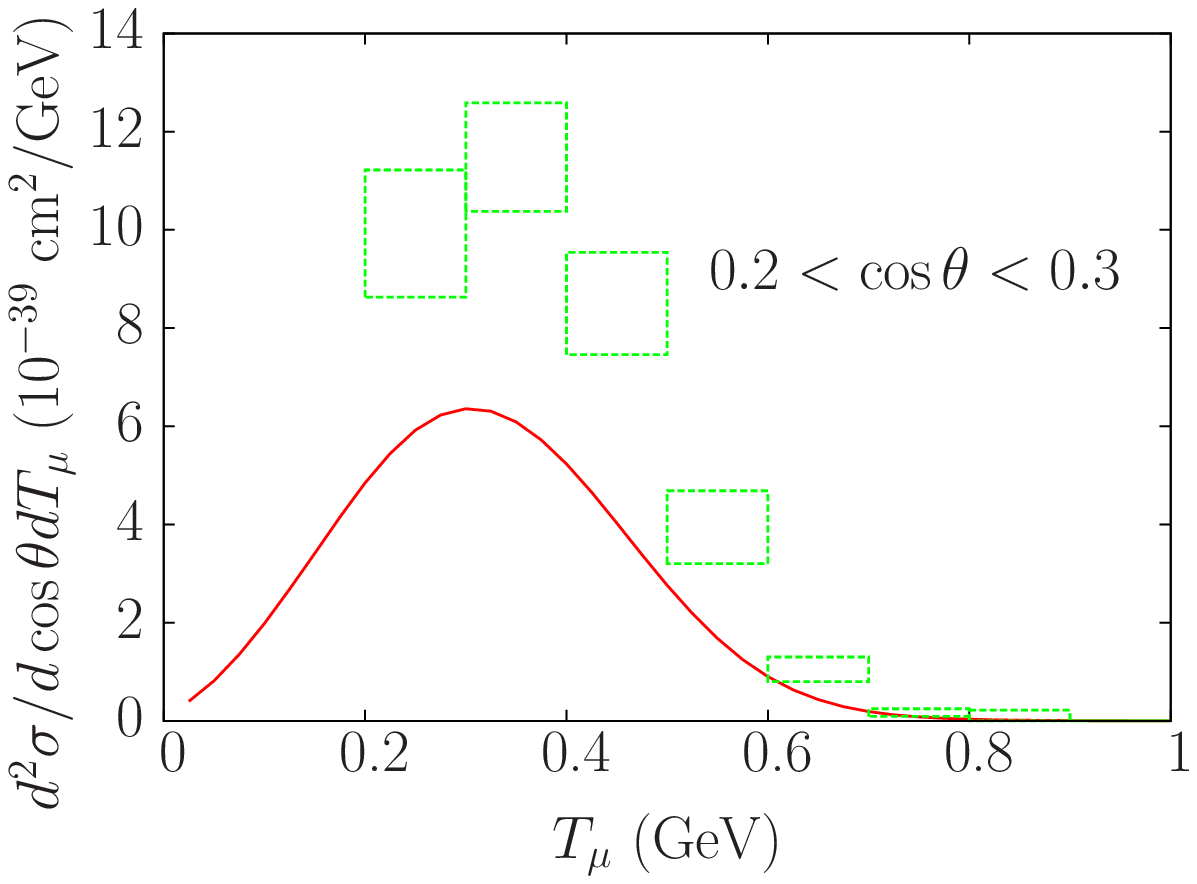}%
\\
\includegraphics[scale=0.35]{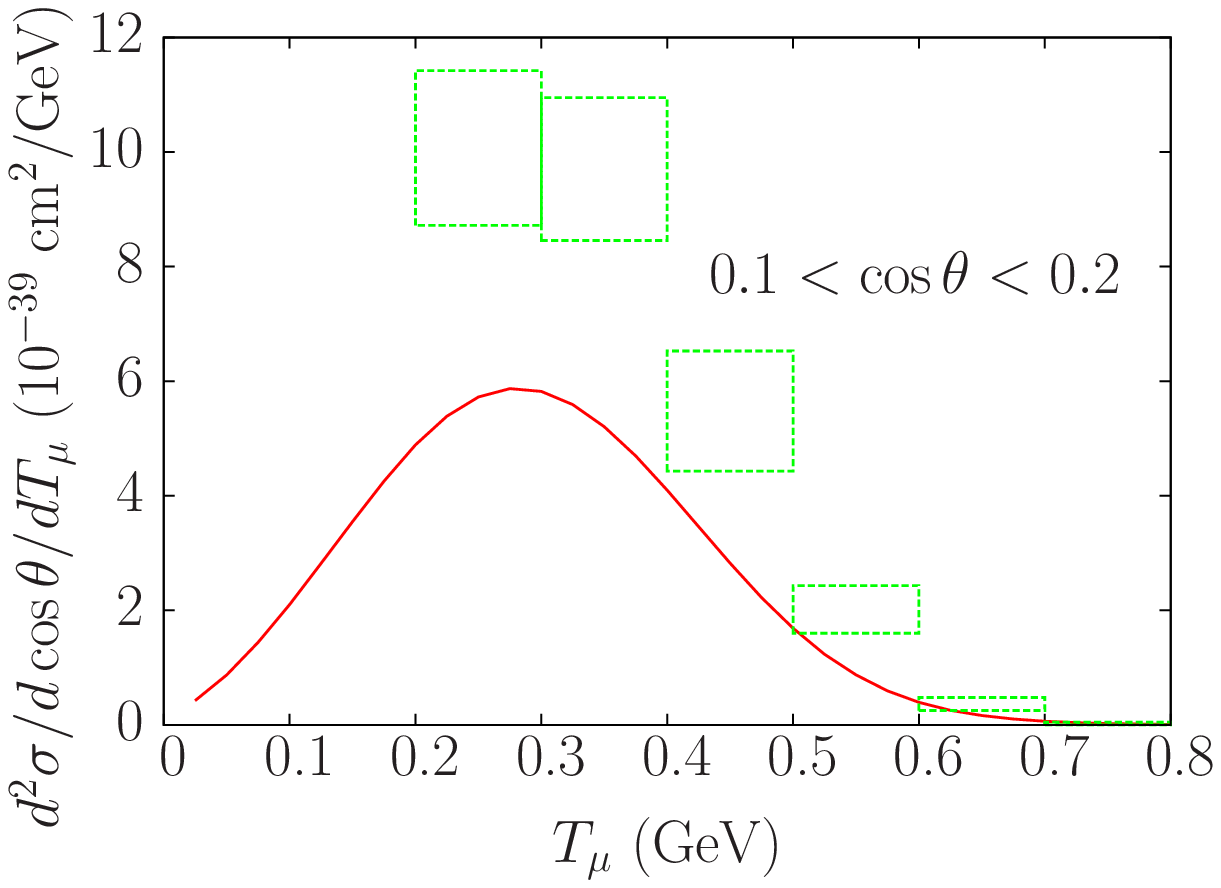}%
\includegraphics[scale=0.35]{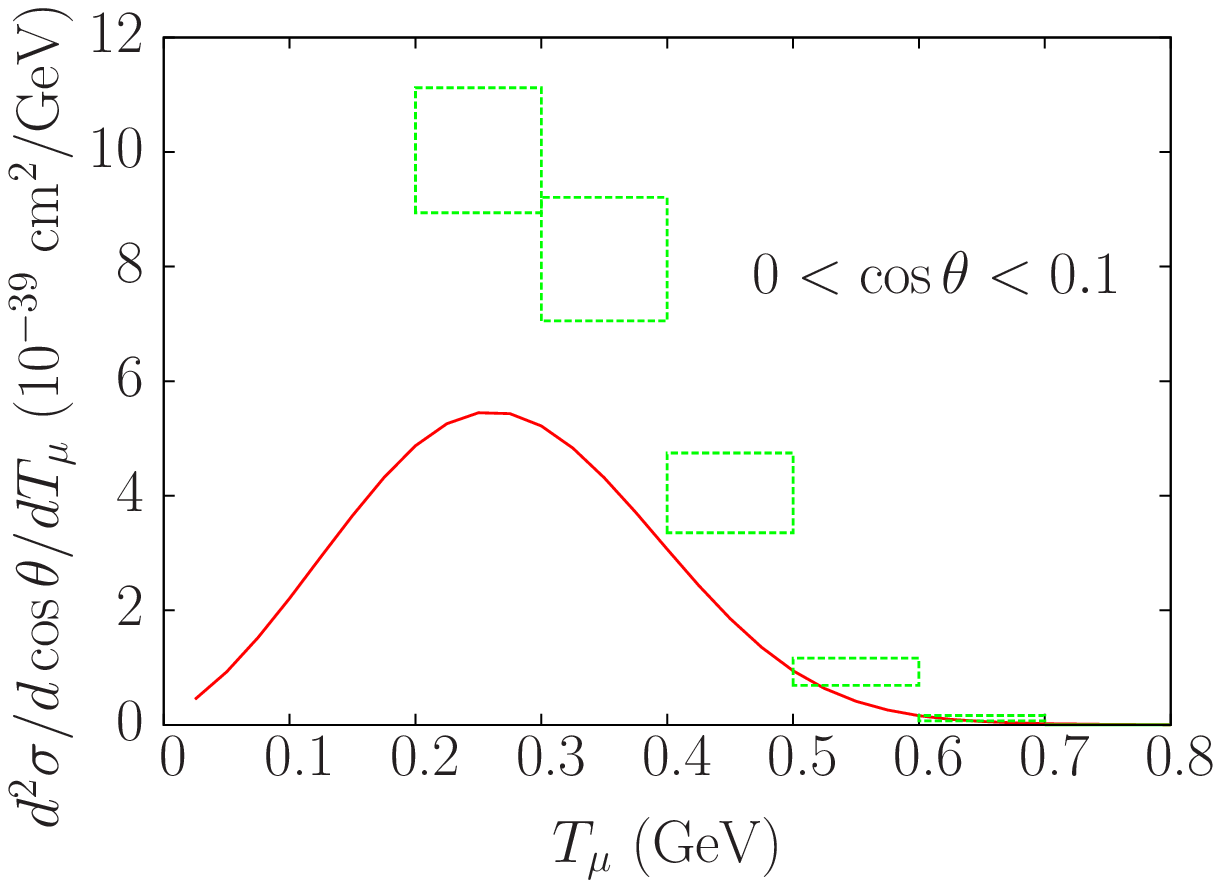}%
\caption{(color online) Flux-integrated double differential cross section per
target nucleon for the $\nu_\mu$ CCQE process on $^{12}$C avaluated in the
SuSA model and displayed
versus the muon kinetic energy $T_\mu$ for various bins of
$\cos\theta$.
The data are from MiniBooNE~\cite{AguilarArevalo:2010zc}. 
The uncertainties do not include
the overall normalization error $\delta N$=10.7\%. }
\end{figure*}
Following \cite{Amaro:2004bs} we write the double differential CCQE
neutrino cross section with respect to the muon scattering angle
$\theta$ and kinetic energy $T_\mu$, at fixed neutrino energy
$E_\nu$, as
\begin{equation}
\left[\frac{d^2\sigma}{d T_\mu d\cos\theta}\right]_{E_\nu}
=\frac{(G_F \cos\theta_{_C})^2 k_\mu}{\pi}
\left(E_\mu-\frac{|Q^2|}{4 E_\nu}\right)
\,{\cal F}^2~,
\label{eq:ddiff}
\end{equation}
where $G_F$ is the Fermi constant,
$\theta_{_C}$  the Cabibbo angle, $k_\mu$  the muon momentum,
$Q^2 = \omega^2-q^2$  the four-momentum transfer, with $\omega=E_\nu-E_\mu$
and ${\vec q}={\vec k_\nu} - {\vec k_\mu}$, and
\begin{eqnarray}
\nonumber
{\cal F}^2 &=& {\hat V}_L R_L^{VV}+
{\hat V}_{CC} R_{CC}^{AA} + 2 {\hat V}_{CL} R_{CL}^{AA} + {\hat V}_{LL} R_{LL}^{AA}
\\
&+&{\hat V}_T \left(R_T^{VV}+R_T^{AA}\right)+
2 {\hat V}_{T^\prime} R_{T^\prime}^{VA}
\end{eqnarray}
the nuclear response.
The kinematical factors ${\hat V}_i$ are given in \cite{Amaro:2004bs}; the nuclear response functions can be cast as
\begin{equation}
R_i = \frac{m_N}{q k_F} R_i^{s.n.} f(\psi) ~,
\end{equation}
where $m_N$ is the nucleon mass, $k_F$ is the Fermi momentum,
$R_i^{s.n.}$ are the single nucleon responses, $\psi(q,\omega)$ is
the RFG scaling variable (see, {\it e.g.}, \cite{Alberico:1988bv}
for its definition) and $f(\psi)$ is the so-called superscaling
function, containing the dependence on the nuclear model. In the RFG
model the latter is a parabola limited to the region $-1<\psi<1$:
$f(\psi)=\frac{3}{4}(1-\psi^2)\theta(1-\psi^2)$. In the SuSA
approach, as already mentioned, it is given by a fit to the
experimental longitudinal $(e,e^\prime)$ reduced response
function~\cite{Jourdan:1996ut}.

In order to compare with the MiniBooNE data we average the cross
section (\ref{eq:ddiff}) over the neutrino energy
flux~\cite{AguilarArevalo:2010zc}. We use the Hoehler
parametrization of the electromagnetic form factors and the value
$M_A=1.03$ GeV/c$^2$ for the nucleon axial mass.
In Fig.~1 we display the flux-integrated double differential cross section
obtained within the SuSA model at fixed $\theta$ as a function of the
muon kinetic energy and compare with the MiniBooNE data~\cite{AguilarArevalo:2010zc}.
Since the experimental data are given in bins of
$\cos\theta$, we display the results averaged over each angular bin.

For most of the angle bins one sees that the SuSA results fall below the data and 
only for low scattering angles (for instance, look at the lowest two angle bins, 
0.9$<\cos\theta<$1 and 0.8$<\cos\theta<$0.9) is the agreement reasonably 
good  (note that an overall normalization error $\delta N$=10.7\%
should also be taken into account). However, one should be very cautious in applying models that are 
devised to work for quasi-free scattering, specifically the RFG model or the present SuSA approach. Such 
models are not well suited to explaining the low-lying excitations in nuclei which arise from discrete states, giant resonances,  { \it  etc.} Indeed, when effects that fall under the heading ``Pauli blocking'' are tested in electron scattering for excitations near threshold one does not see good agreement between experiment and such simple modeling. The RFG model, the SuSA approach or any other models that lack the ability to address the complexity of the many-body problem in the near-threshold region are not supposed to be applied in this low $q$--$\omega$ regime. A proper treatment (for instance, using RPA with realistic nuclear
wave functions) of
collective excitations is clearly required. Accordingly it is important to analyze how much of the integrated strength in the various panels in Fig.~1 arises from the low excitation region and how much from larger energies where the modeling may be more robust.

To make such an assessment, in Fig.~2 we show the SuSA results obtained
by integrating over the full neutrino flux (solid lines, red online) and by artificially
cutting the integral at energy transfer $\omega$=50 MeV (dashed lines, green online). It clearly appears
that at the most forward angles (upper left panel) when
the results are cut at 50 MeV the cross section drops by about a
factor of 2, showing that roughly 1/2 of the cross section for such
kinematics arises from the first 50 MeV of excitation. For 
more backward angles (upper right and lower panels) the cut effect is much weaker, 
indicating that low excitation energies are only significant for the 
first angle bin. We are thus forced to conclude that approaches such as SuSA in the present paper (or the RFG, even when Pauli blocking effects are incorporated) should not be trusted for these very forward, significantly low-energy kinematics.

\begin{figure}[ht]
\label{fig:cut}
\includegraphics[scale=0.35]{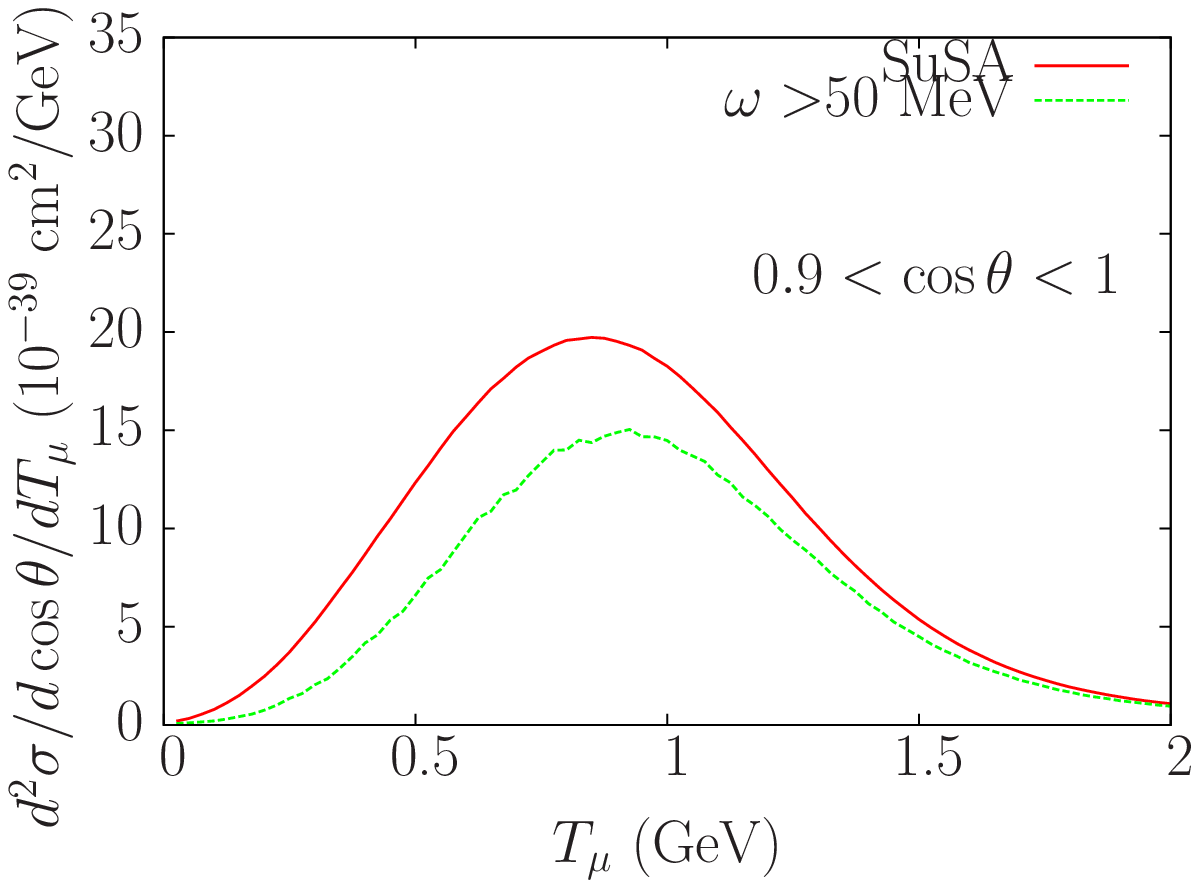}%
\includegraphics[scale=0.35]{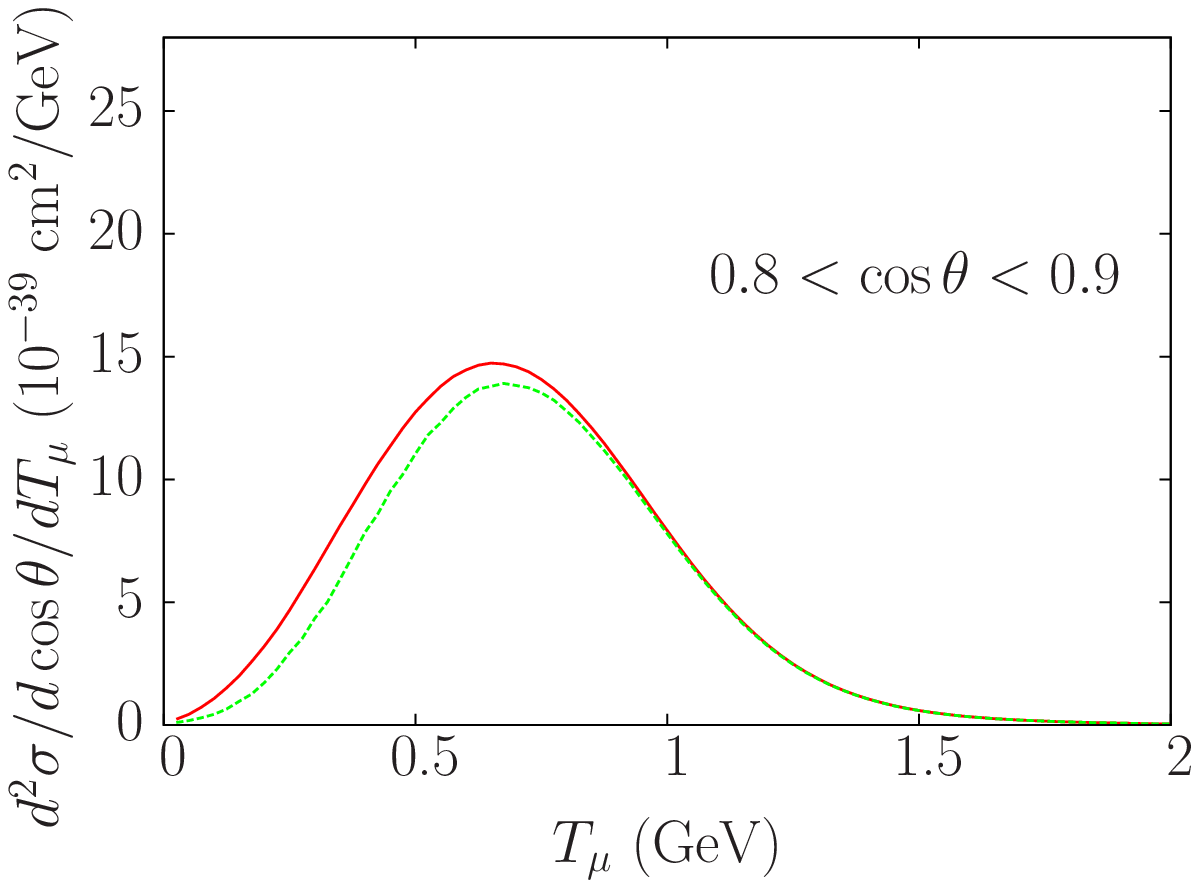}\\
\includegraphics[scale=0.35]{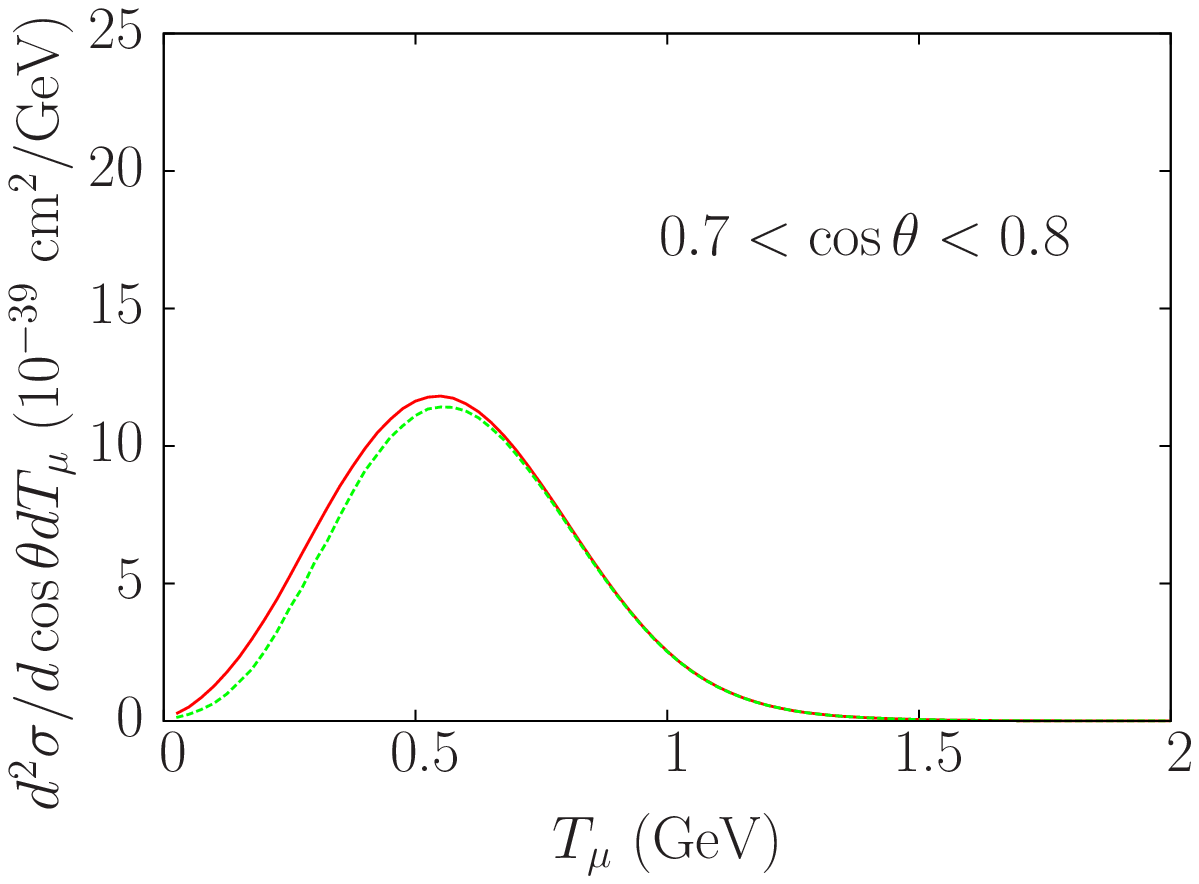}%
\caption{(color online) solid lines (red online): flux-integrated cross sections calculated in the SuSA 
model for specific bins bin of scattering angles. dashed lines (green online): a lower
cut $\omega=50$ MeV is set in the integral over the neutrino flux.}
\end{figure}
In passing one should note that an additional correction arises from the distortion of the outgoing muon wave function in the
Coulomb field of the recoiling nucleus. A rigorous description of this effect is somewhat 
complicated~\cite{Coulomb}, as for the outgoing lepton it requires the use of distorted waves
which are eigenfunctions of the nuclear Coulomb field; however its main effects can be described
using approximate approaches. In particular, the effective momentum 
approximation~\cite{Alberico:1997vh,Amaro:2004bs,Maieron:2003df}
has been successfully applied to the case of medium-to-high energy leptons
and is employed here. For muon kinetic energies
$T_\mu\geq 200$ MeV, as considered in the MiniBooNE experiment, Coulomb effects are
below $\sim 2\%$. Obviously, for smaller $T_\mu$-values Coulomb distortion produces bigger effects, these
being on the order of $\sim 10\%$ for $T_\mu\sim 90-100$ MeV (see \cite{Alberico:1997vh}).

\section{Two-particle two-hole meson-exchange currents}

\begin{figure*}[ht]
\label{fig:cosMEC}
\includegraphics[scale=0.35]{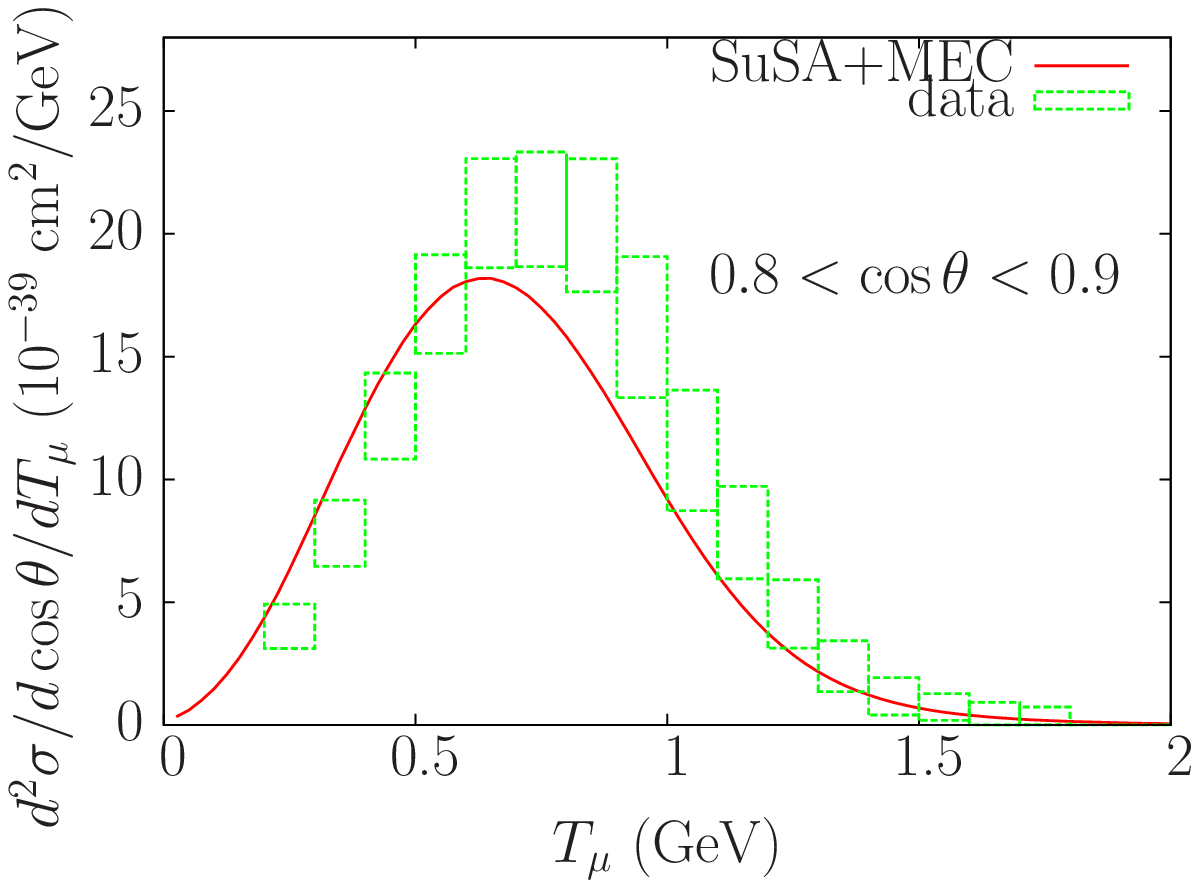}%
\includegraphics[scale=0.35]{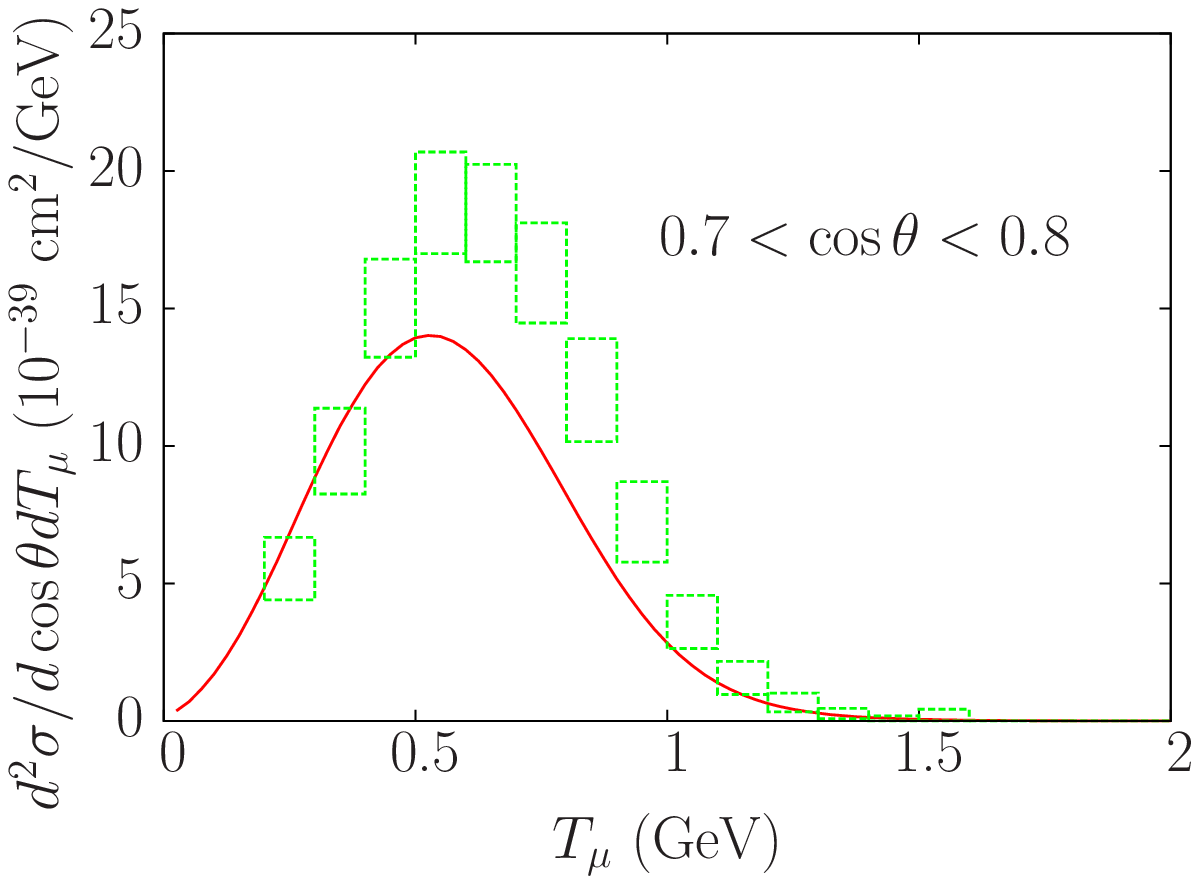}%
\includegraphics[scale=0.35]{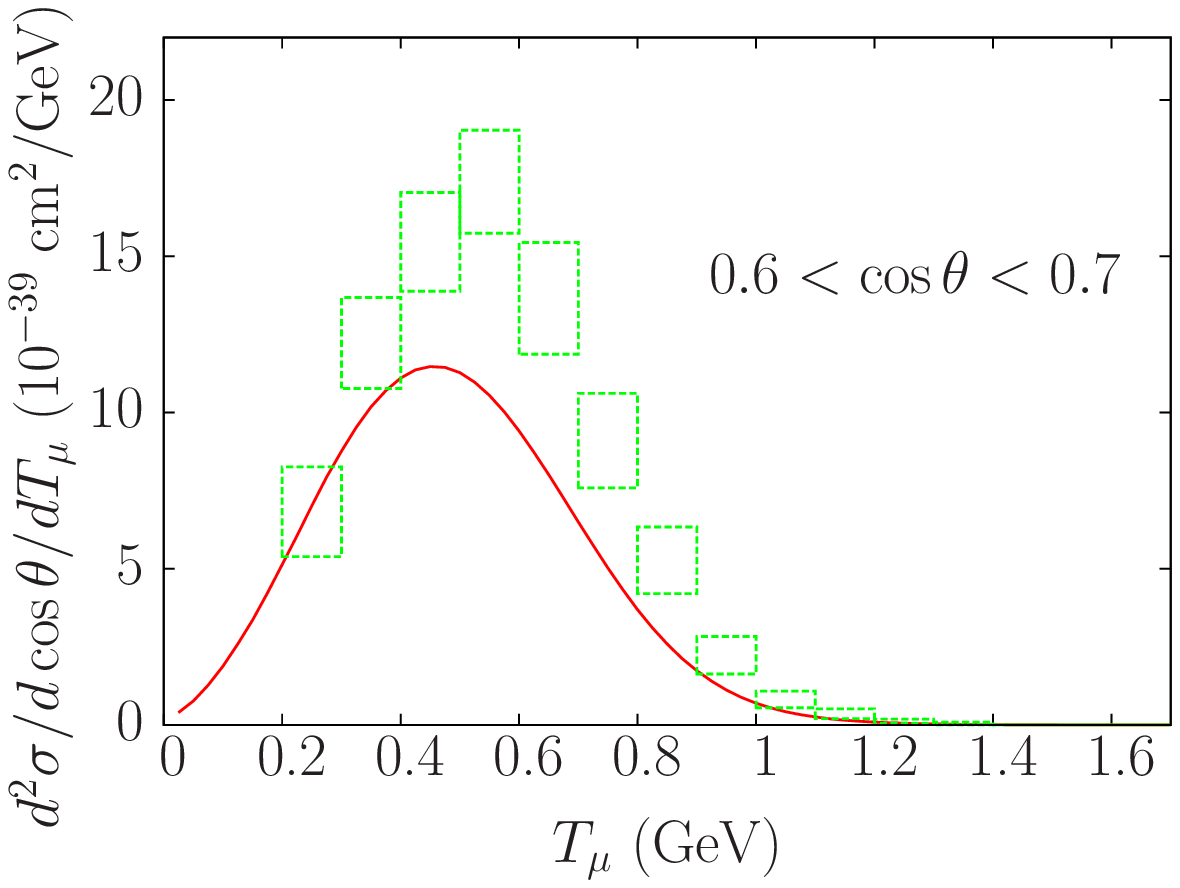}%
\\
\includegraphics[scale=0.35]{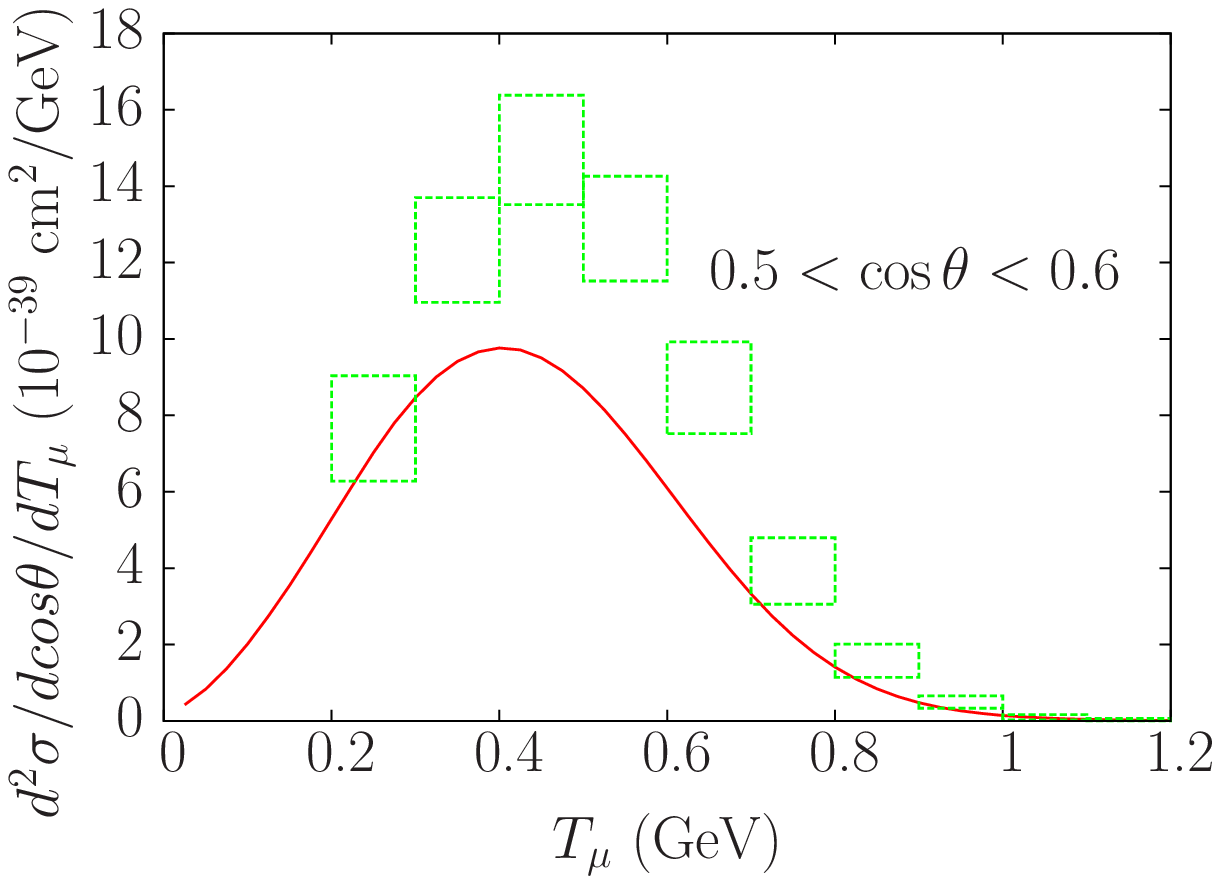}%
\includegraphics[scale=0.35]{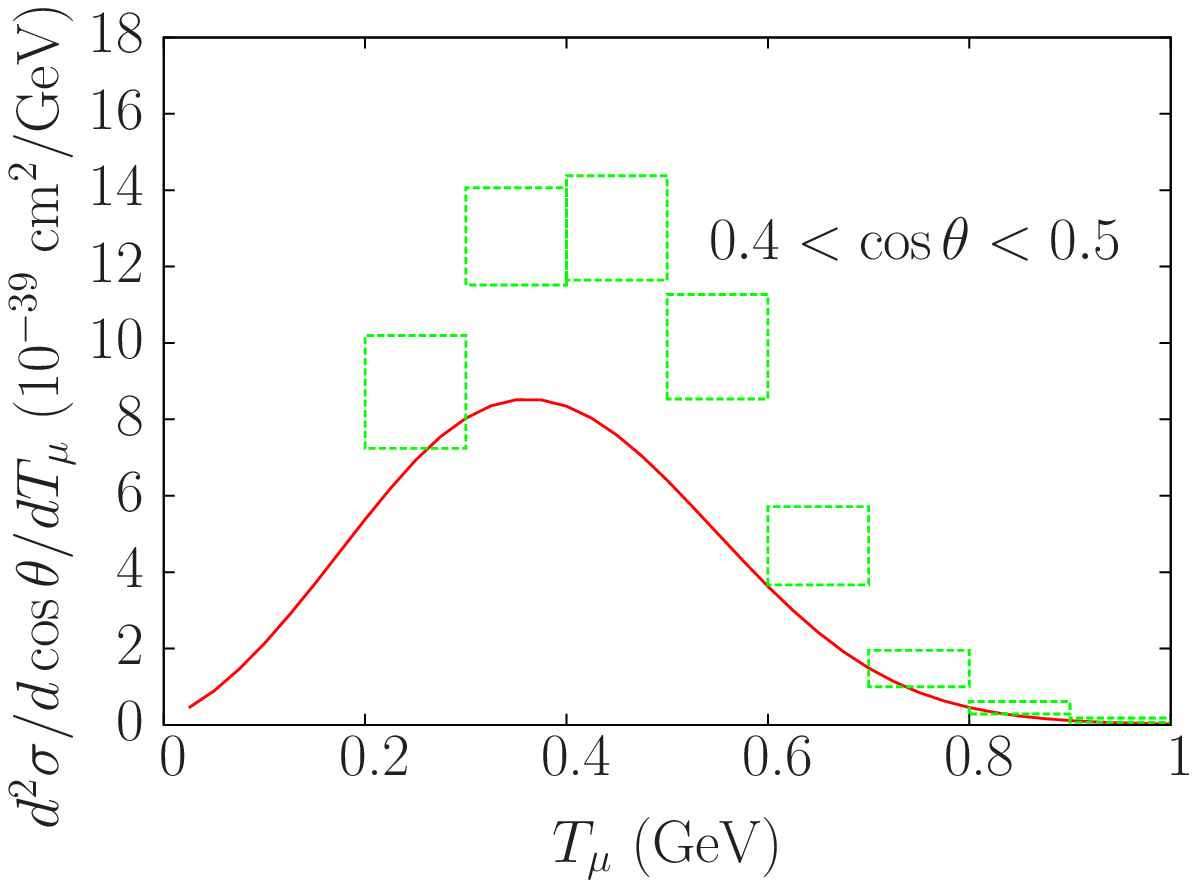}%
\includegraphics[scale=0.35]{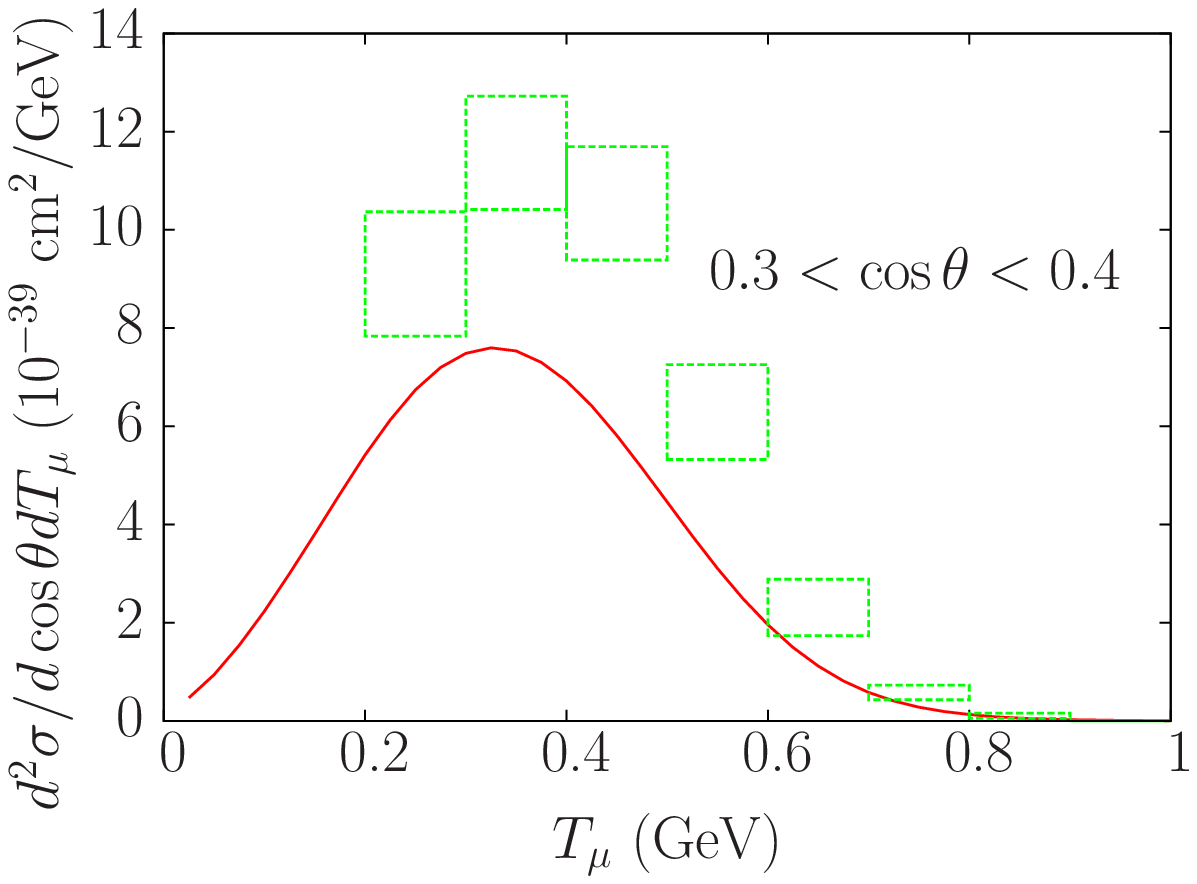}%
\\
\includegraphics[scale=0.35]{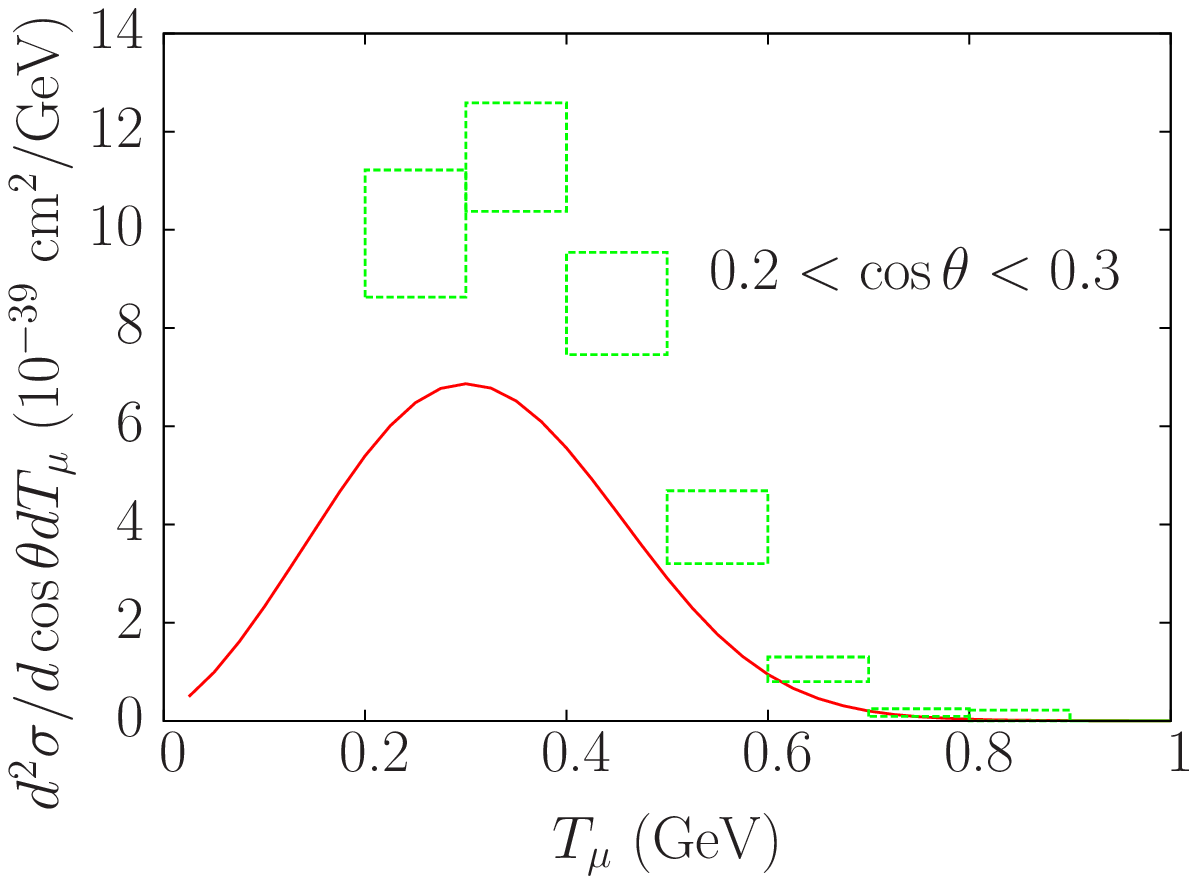}%
\includegraphics[scale=0.35]{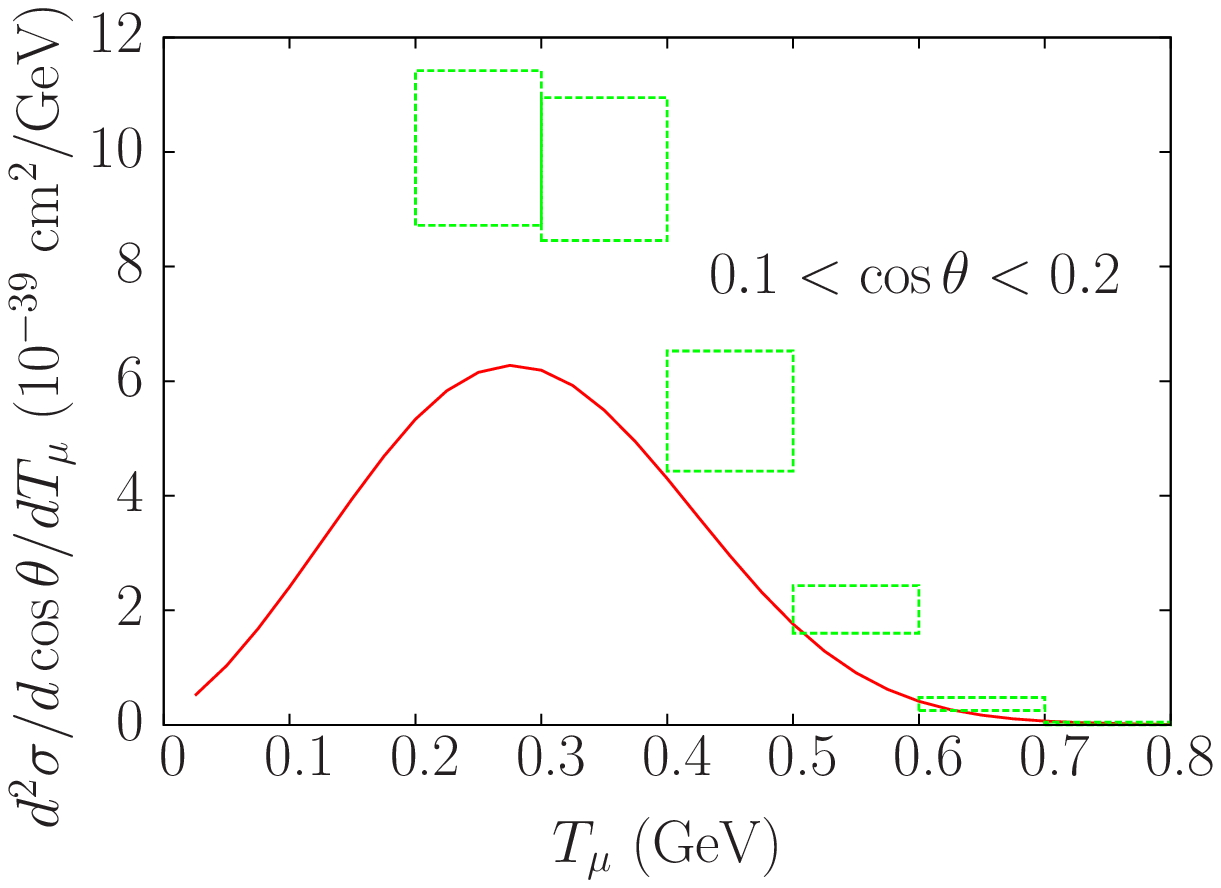}%
\includegraphics[scale=0.35]{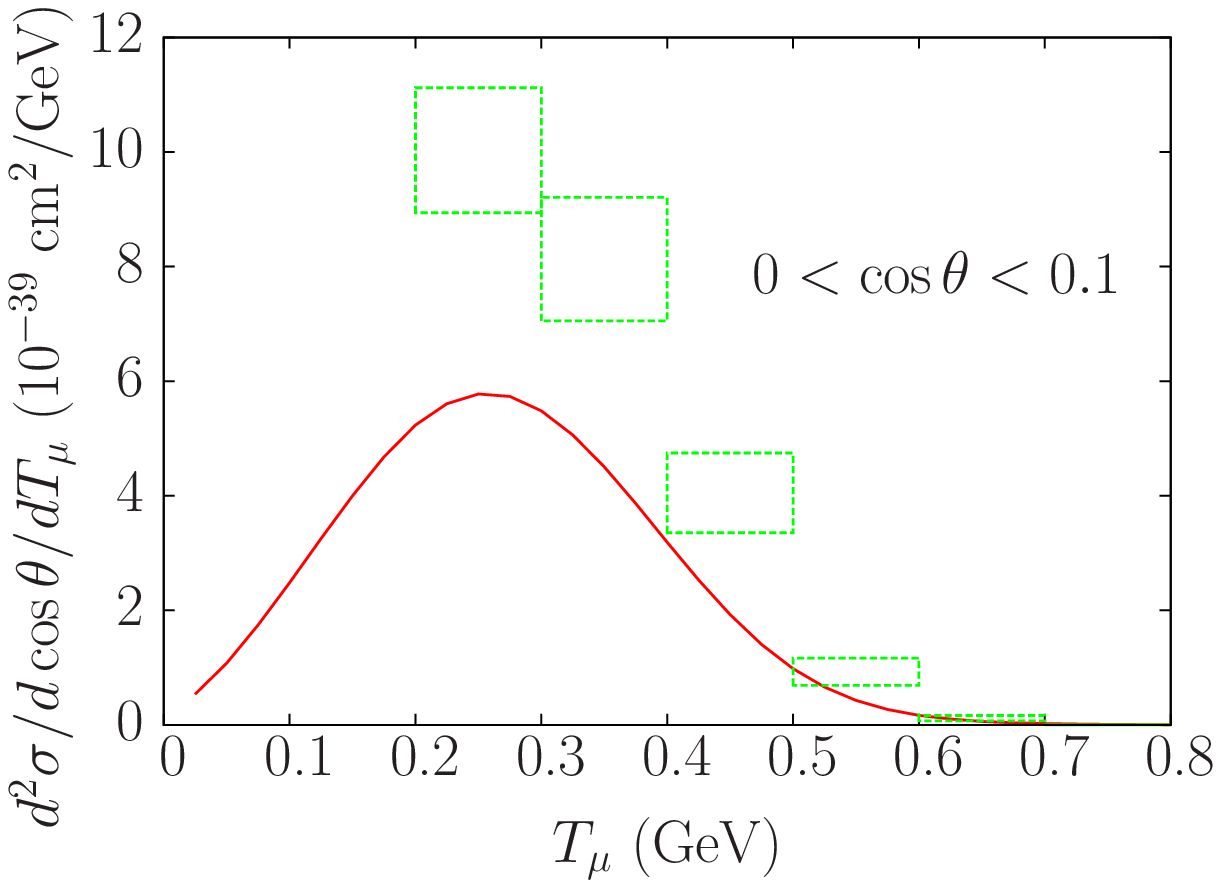}%
\caption{(color online) Same as Fig.~1 for the SuSA model, but now including 2p2h
meson-exchange currents.}
\end{figure*}
In this section we evaluate the contribution of meson-exchange
currents to the CCQE cross section. The MEC are two-body currents,
and therefore can excite both one-particle one-hole (1p-1h) and
two-particle two-hole (2p-2h) final states.

Most of MEC studies of electromagnetic (e,e$^\prime$) processes
performed for low-to-intermediate momentum transfers in the 1p-1h
sector (see, {\it e.g.},
\cite{Alberico:1989aja,Amaro:2002mj,Amaro:2003yd,Amaro:2009dd}) have shown a
small reduction of the total response at the quasielastic peak,
mainly due to diagrams involving the electroexcitation of the
$\Delta$ resonance. These roughly compensate the positive
contribution due to correlation diagrams, where the virtual photon
couples to a correlated pair of nucleons. In the present work we
shall therefore neglect them and restrict our attention to 2p-2h
final states.

The impact of pionic 2p-2h MEC on inclusive electron scattering
reactions has first been evaluated in the RFG framework in
\cite{VanOrden:1980tg}, where a non-relativistic reduction of the
currents was performed. Fully relativistic calculations have been
developed more recently in \cite{Dekker:1994yc,De Pace:2003xu}. It
has been found that the MEC give a significant positive contribution
to the cross section, which helps to account for the discrepancy
between theory and experiment in the ``dip'' region between the
quasielastic and $\Delta$-resonance region. Moreover, the MEC
have been shown to break scaling of both first and second
kinds~\cite{De Pace:2004cr}.

In this paper we use the fully relativistic model of \cite{De
Pace:2003xu}, where all many-body diagrams containing two pionic
lines that contribute to the electromagnetic 2p-2h transverse response
were taken into account. Note that in lowest order these affect only
the transverse polar vector response, $R_T^{VV}$.
\begin{figure*}[ht]
\label{fig:tmu}
\includegraphics[scale=0.35]{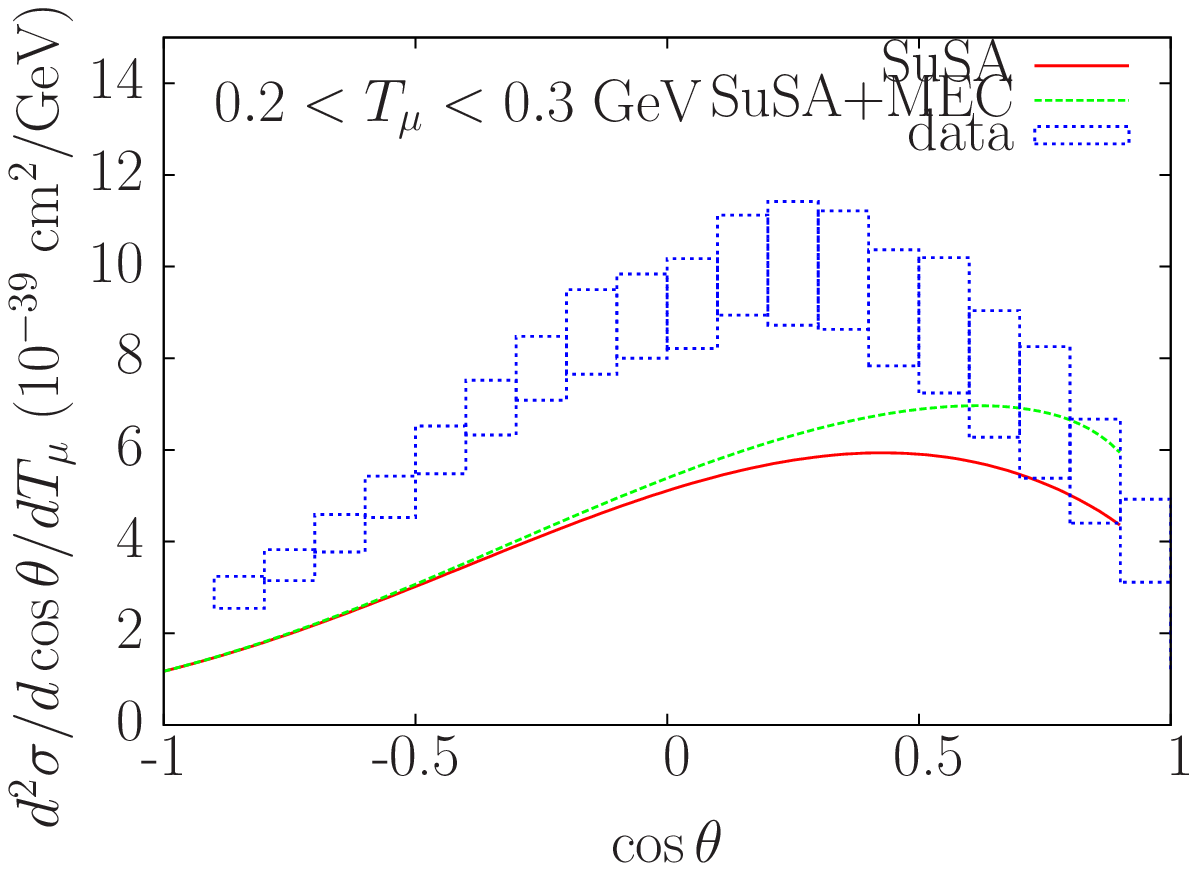}%
\includegraphics[scale=0.35]{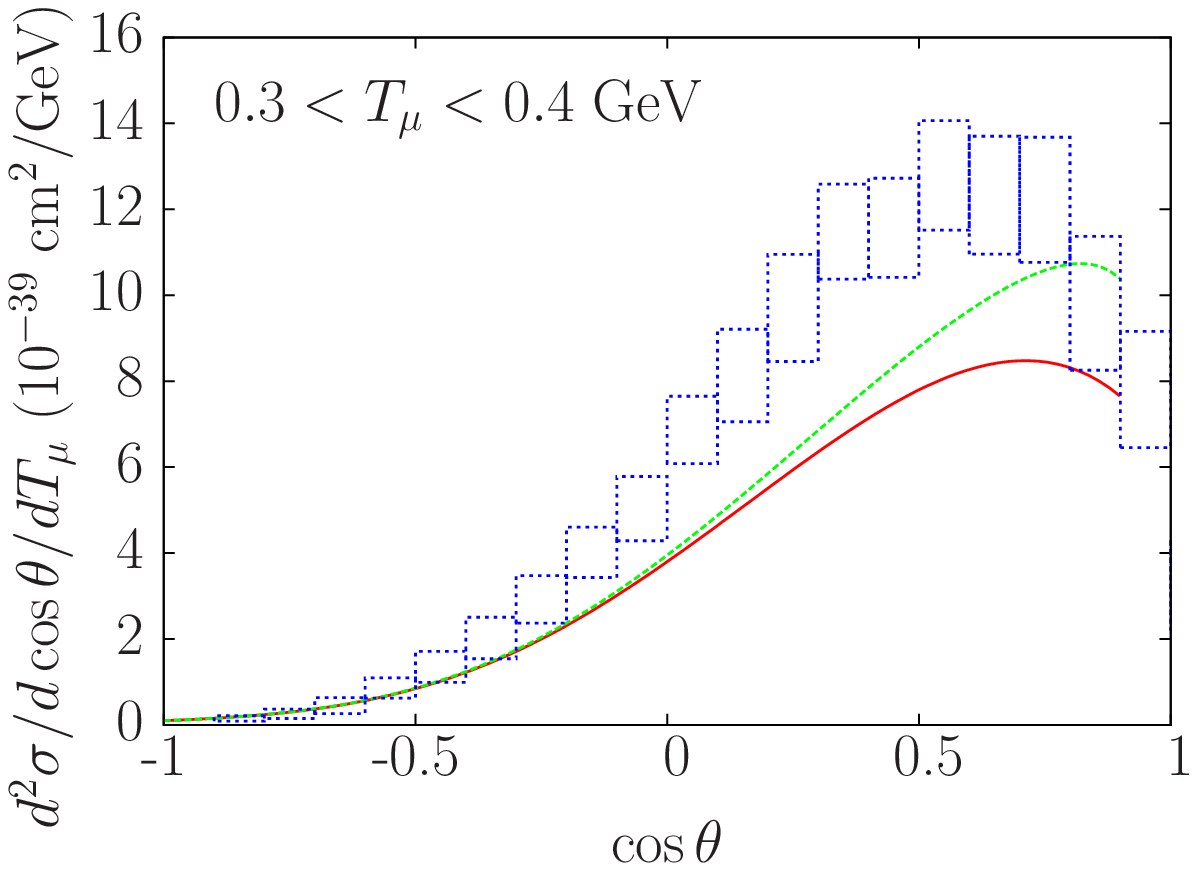}%
\includegraphics[scale=0.35]{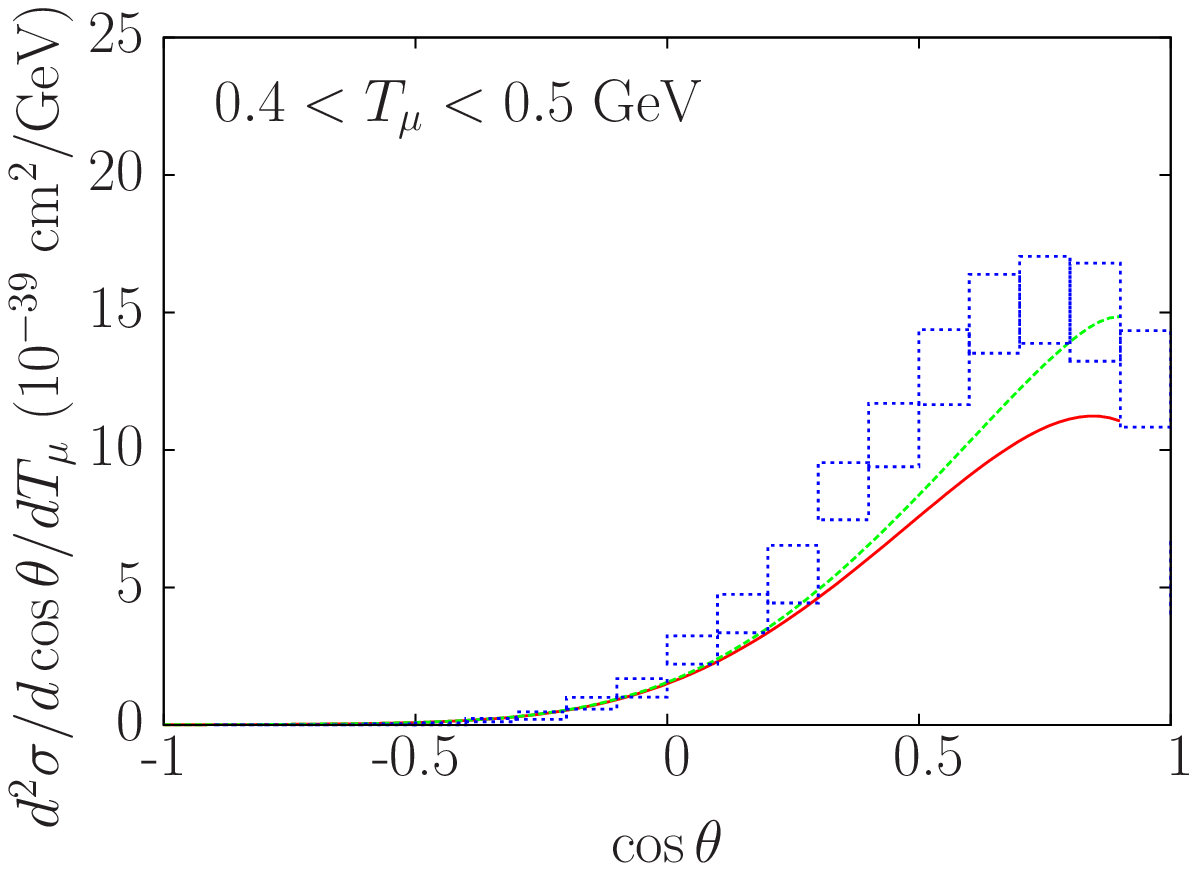}%
\includegraphics[scale=0.35]{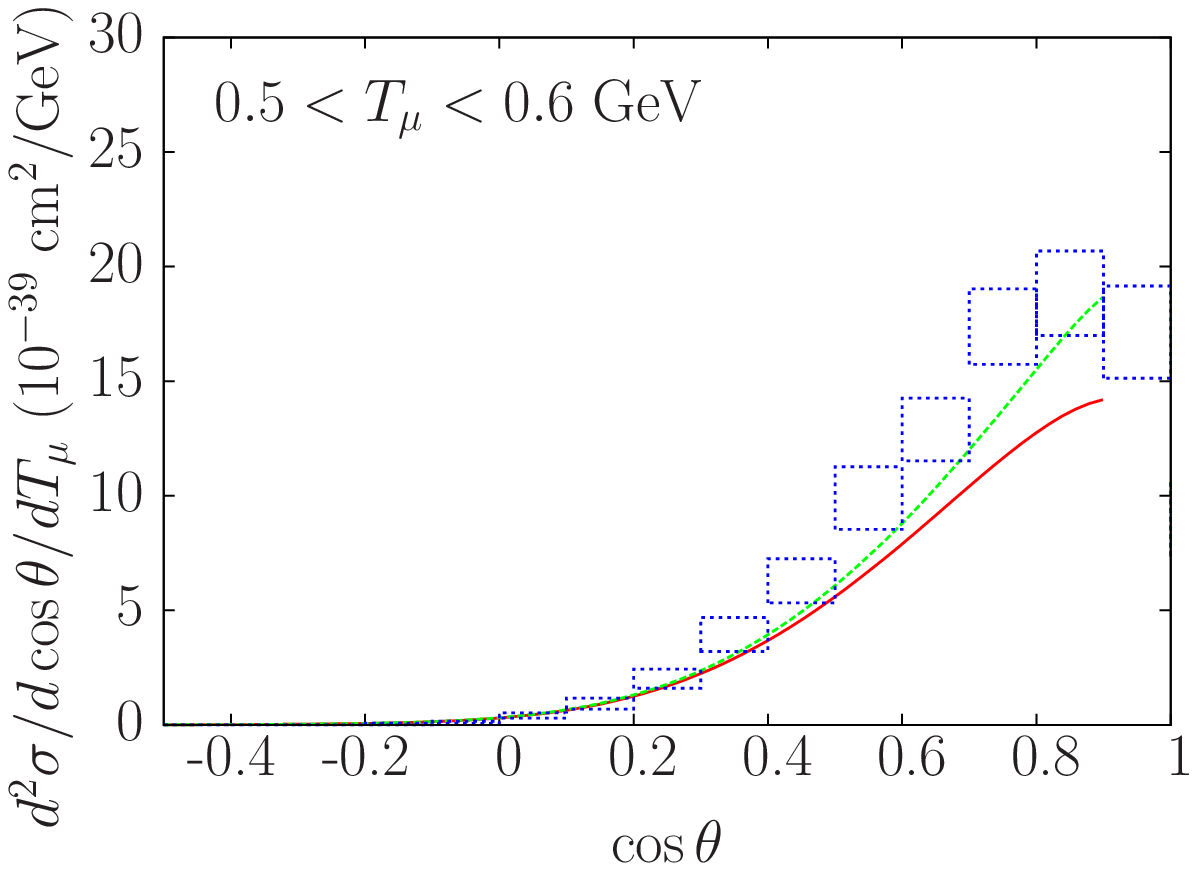}%
\\
\includegraphics[scale=0.35]{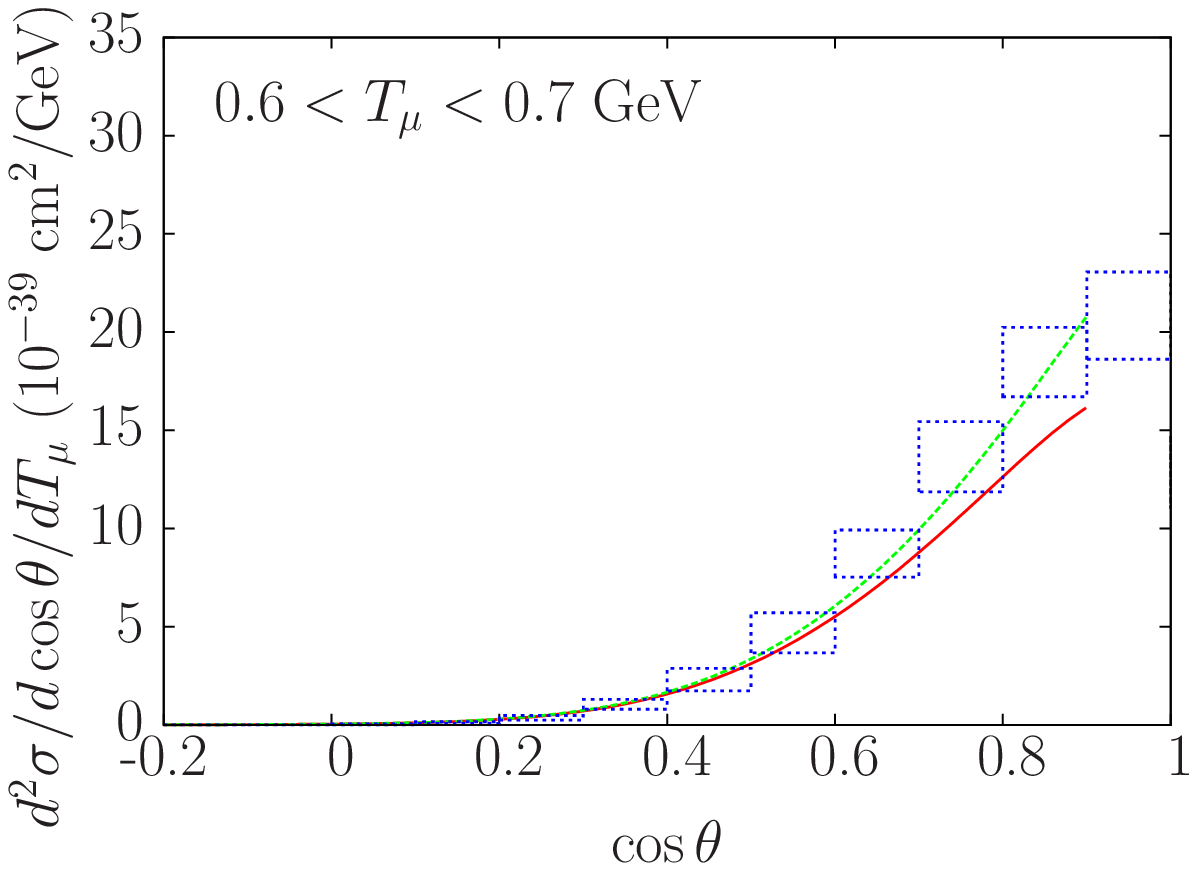}%
\includegraphics[scale=0.35]{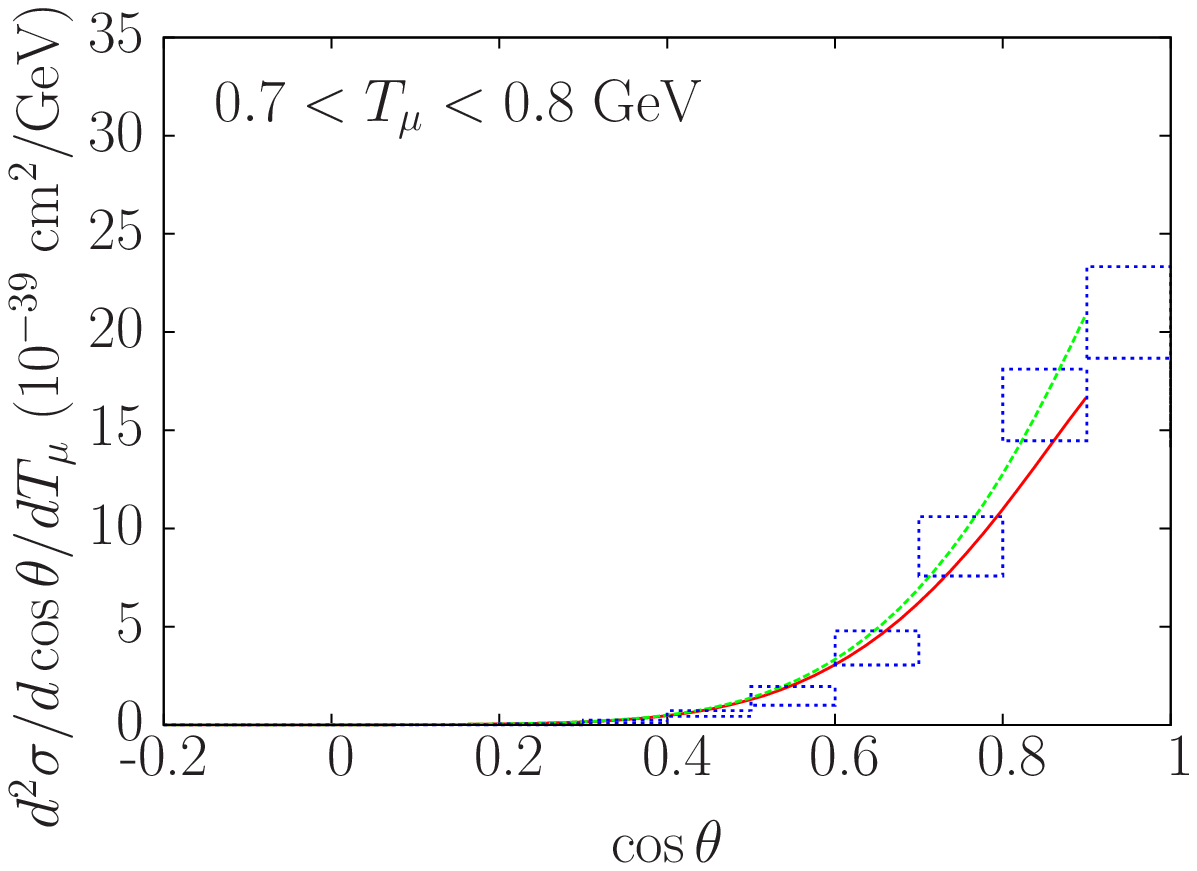}%
\includegraphics[scale=0.35]{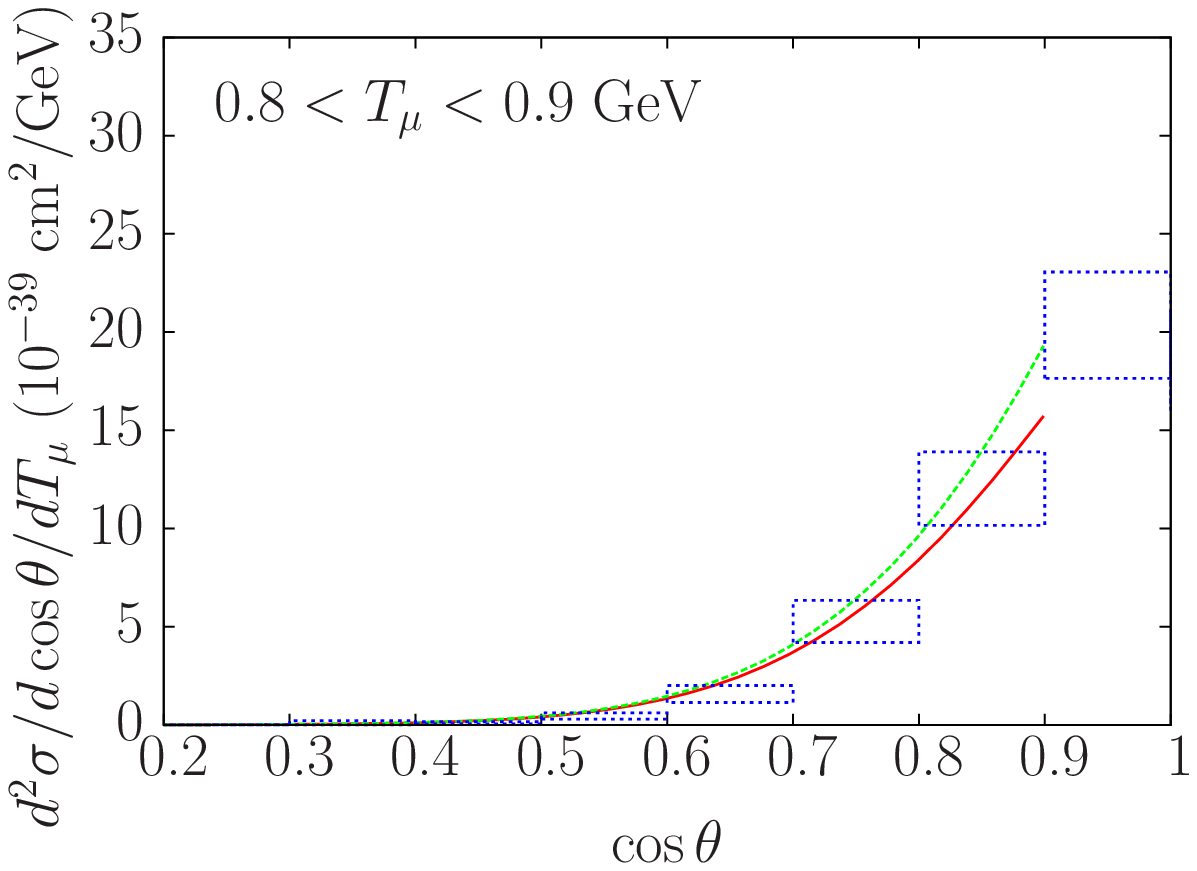}%
\includegraphics[scale=0.35]{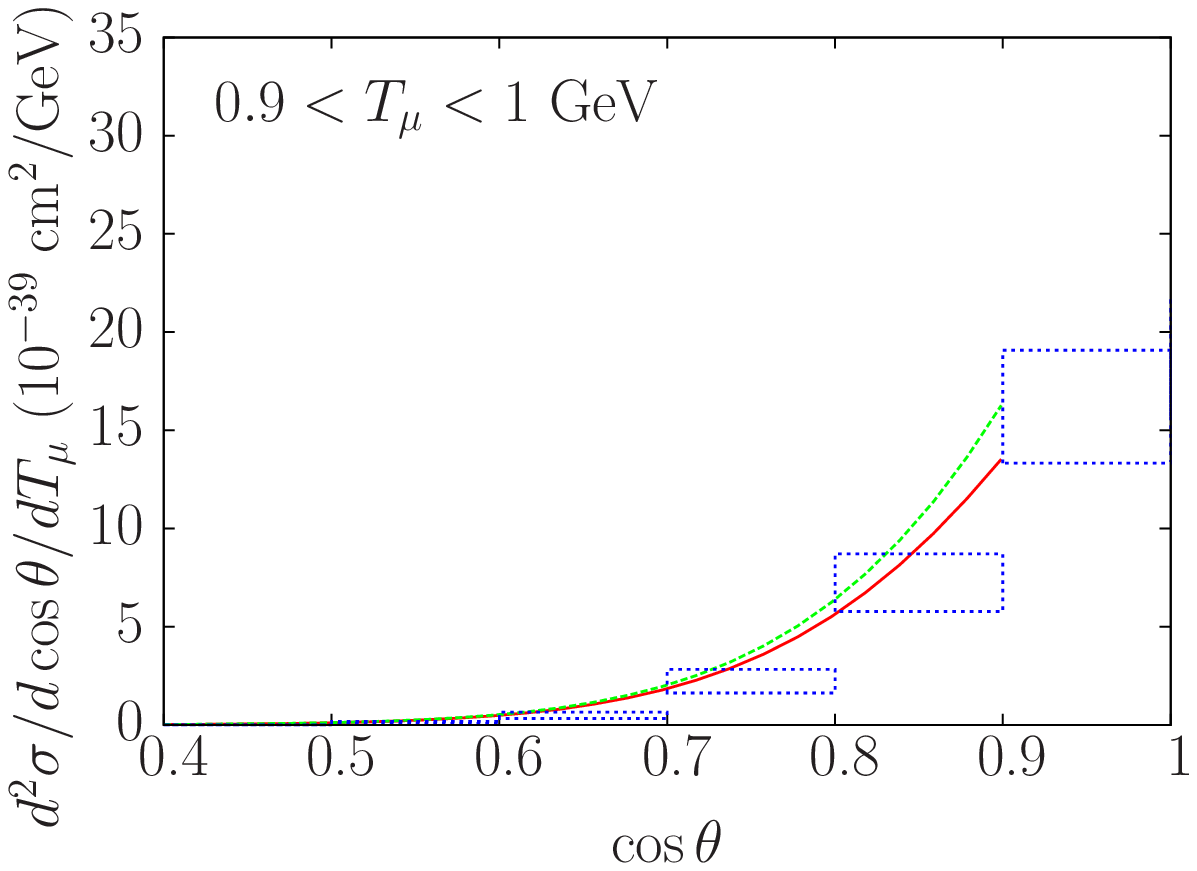}%
\caption{(color online) Double differential $\nu_\mu$ CCQE cross section for
$^{12}$C integrated over neutrino flux versus the outgoing muon
scattering angle for various bins of the muon kinetic energy
$T_\mu$. Solid lines (red online): SuSA; dashed lines (green online): 
SuSA model including the MEC 2p-2h contribution. In the light of the discussion of Fig.~2 given above, results are given only for the region $\cos\theta<$0.9.}
\end{figure*}
As shown in Fig.~3, the 2p-2h MEC tend to increase the cross
section, yielding reasonable agreement with the data out to
$\cos\theta\simeq 0.6$. At larger angles the disagreement with
the experiment becomes more and more significant, and the meson-exchange currents are not sufficient to account for the discrepancy. We also note that recent studies using somewhat different assumptions~\cite{Amaro:2p2h} yield similar results for 2p-2h MEC contributions, although these tend to be somewhat larger than the earlier results employed in the present work, and would lead to somewhat better agreement with the CCQE data at larger angles.

The same conclusion discussed above can be drawn by plotting the cross section
versus the scattering angle at fixed $T_\mu$ (Fig.~4): the inclusion
of two-body currents improves the agreement with the data at low
scattering angles, but some strength is missing at higher angles,
especially for low muon momenta.

Again some caution should be expressed before drawing definitive conclusions from the agreements or disagreements seen in Figs.~3 and 4. For instance, as already mentioned, there are strong indications from RMF studies as well as from QE $(e,e')$ data that scaling of the zeroth kind is only approximate and that the vector transverse response should be enhanced over the strict SuSA strategy employed in the present work. Moreover, in different language, namely that of extended RFG modeling where 1p-1h and 2p-2h excitations are incorporated, a fully consistent treatment should take
into account not only the MEC contributions of the present study but also the correlation diagrams that are 
necessary in order to preserve the gauge invariance of the
theory. In an infinite system like the RFG these diagrams
give rise to divergencies which need to be regularized. A treatment of these
contributions has been performed recently for $(e,e')$ in
\cite{Amaro:2p2h}, where they were shown to be of the same order as
the MEC.  The model of \cite{Amaro:2p2h} contains similar ingredients
in the treatment of 2p-2h excitations to those in \cite{Mar09,Mar10} with
the addition of being fully relativistic. These contributions also enhance the cross sections beyond the results shown above and so might be responsible for the residual disagreement. Indeed the trend is encouraging: their effect grows with increasing momentum transfer and hence they have a greater impact for large neutrino-muon angles than for the forward direction where the results shown in Figs.~3 and 4 are already reasonably successful. It is too early to say for sure, however, since there is at present no completely consistent relativistic model that is capable of incorporating all of the effects discussed above.
\section{Conclusions}
In summary, we have applied the phenomenological model based on
electron scattering data, elaborated in \cite{Amaro:2004bs}, to CCQE
neutrino reactions and compared the results with the recent
MiniBooNE double differential quasielastic cross section data at all
available kinematics. In addition to presenting detailed results for the so-called SuSA approach, two specific issues have been addressed in the present work: (1) the role played by 2p-2h MEC contributions has now been explored, and (2) cross sections at small angles have clearly been shown to be related to the regime where low-energy nuclear excitations dominate and thus where quasi-free modeling must be viewed with suspicion. The strict SuSA predictions show a systematic discrepancy
between data and theory, where they tend to underestimate the data
especially at large muon scattering angles and low muon energies. When 2p-2h MEC contributions are included the situation is different: inclusion of the 2p-2h contributions yields results that are compatible with the data for
$\theta \leq 50^0$ (excluding the most forward angles where quasi-free modeling must be questioned, as stated above), but lie below the data at larger angles where the predicted cross sections are smaller. These two-body currents arise from
microscopic relativistic modeling performed for inclusive electron
scattering reactions and they are known to result in a significant
increase in the vector-vector transverse response function, in concert with QE electron scattering data. It should,
however, be remembered that the present approach still lacks the
contributions from the correlation diagrams associated with the MEC which
are required by gauge invariance; these might improve the agreement
with the data, as suggested by the results of \cite{Amaro:2p2h} for
inclusive electron scattering. Finally, we note that alternative approaches such as relativistic mean field theory also lead us to expect an enhancement over the results shown in the present work, as, in fact, are observed for QE electron scattering data in a similar kinematic region. Work is in progress to resolve several of these issues, specifically, to perform a detailed study of modeling versus experiment for inclusive QE electron scattering, to extend the analysis based on RMF both for electron scattering and for neutrino reactions, and to incorporate the missing pieces mentioned above that are required to restore gauge invariance. Once these are in hand it will be appropriate to re-visit the issue of the anomalous axial mass.

%
\section*{Acknowledgments}
We thank Arturo De Pace for providing the calculation of the
electromagnetic 2p-2h responses. This work was
partially supported by DGI (Spain): FIS2008-01143,
FIS2008-04189, by the Junta de
Andaluc\'{\i}a, by the INFN-MEC collaboration agreement, projects
FPA2008-03770-E-INFN, ACI2009-1053, the Spanish Consolider-Ingenio
2000 programmed CPAN (CSD2007-00042), and part (TWD) by U.S.
Department of Energy under cooperative agreement DE-FC02-94ER40818.

%

\end{document}